\documentclass[a4paper,11pt]{article}
\usepackage{jcappub}
\usepackage[utf8]{inputenc}
\usepackage{amsmath, amssymb, amsthm, graphicx, epsfig, fancyhdr,epsfig, slashed}
\usepackage{xcolor}
\usepackage{color}
\usepackage{mathrsfs}
\usepackage{graphicx}
\usepackage{dcolumn}
\usepackage[utf8]{inputenc}
\usepackage[normalem]{ulem}
\DeclareUnicodeCharacter{2212}{-}
\usepackage{bm}
\usepackage{hyperref}
\usepackage[mathlines]{lineno}
\usepackage{hyperref}
\usepackage[all]{hypcap}
\hypersetup{  
    colorlinks=true,
    linkcolor=blue,
    filecolor=red,      
    urlcolor=cyan,
    citecolor=green,
    }
\newcommand{\be}{\begin{equation}}
\newcommand{\ee}{\end{equation}}
\newcommand{\bea}{\begin{aligned}}
\newcommand{\eea}{\end{aligned}}

\newcommand{\bse}{\begin{subequations}}
\newcommand{\ese}{\end{subequations}}

\newcommand{\bmm}{\begin{multline}}
\newcommand{\emm}{\end{multline}}

\usepackage[dvipsnames]{xcolor}

\title{Precision Analysis for $\boldsymbol{H_0}$ Using Upcoming Multi-band Gravitational Wave Observations}

\author[1]{Setabuddin,}
\emailAdd{setabuddin.ahmad@gmail.com}
\affiliation[1]{Physics and Applied Mathematics Unit, Indian Statistical Institute, 203 B.T. Road, Kolkata 700108, India}
\author[1,2]{Md Riajul Haque,}
\emailAdd{riaj1994sjtu.edu.cn}
\affiliation[2]{Tsung-Dao Lee Institute \& School of Physics and Astronomy, Shanghai Jiao Tong University, Shanghai 201210, China}
\author[3]{Ratna Koley,}
\emailAdd{ratna.physics@presiuniv.ac.in}
\affiliation[3]{Department of Physics, Presidency University, 86/1 College Street,
Kolkata 700073, India}
\author[1]{Supratik Pal}
\emailAdd{supratik@isical.ac.in}

\abstract{
We investigate how multi-band gravitational wave (GW) observations can constrain the uncertainties in the Hubble parameter ($H_0$) using primordial black holes (PBHs) as possible sources. Our framework combines scalar-induced and merger-induced GWs from PBHs, and forecasts on a combination of two future detectors  Square Kilometre Array (SKA) and the Einstein Telescope (ET), enabling a multi-band analysis. We perform a statistical forecast of the PBH parameters, $M_{\rm PBH}$ and $f_{\rm PBH}$, using signal-to-noise ratio (SNR) estimates and Fisher matrix analysis. Imposing $\mathrm{SNR} \geq 1$, we identify the accessible PBH parameter space and propagate these uncertainties to estimate the corresponding uncertainties in $H_0$.
 For $\delta \theta_i/\theta_i \leq 0.1$, with $\theta_i \equiv M_{\rm PBH}(f_{\rm PBH})$, we find 
$\delta H_0 \lesssim 2~{\rm km\,s^{-1}\,Mpc^{-1}}$ in a conservative approach, improving to 
$\delta H_0 \lesssim \mathcal{O}(0.1)~{\rm km\,s^{-1}\,Mpc^{-1}}$ 
for $\delta \theta_i/\theta_i \leq 0.01$ for an optimistic approach of precision measurement. The results are further found to be largely insensitive to the fiducial choice of the $H_0$, with only moderate dependence on the PBH collapse efficiency. These findings demonstrate that multi-band GW observations provide an independent and complementary approach to constraining the uncertainties in $H_0$, with the potential to provide a novel, cosmic distance ladder-independent  measure of the Hubble parameter.
}
\renewcommand{\thesection}{\arabic{section}}

\begin{document}
\maketitle
\flushbottom
\section{Introduction}

The next generation gravitational wave (GW) detectors have ushered in the era of GW astronomy, leading to significant advances in astrophysics, cosmology, and fundamental physics. In cosmology, the joint detection of a GW signal and its electromagnetic counterpart from a neutron star merger~\cite{LIGOScientific:2017vwq, LIGOScientific:2017adf, Schutz:1986gp, LISACosmologyWorkingGroup:2025vdz} enabled the first measurement of the Hubble parameter using GW missions. Beyond the chirp-like signals from compact binary coalescences detected by the LIGO/Virgo/KAGRA (LVK) collaboration \cite{LIGOScientific:2016aoc, LIGOScientific:2017vwq, Abramovici:1992ah}, a stochastic gravitational wave background (SGWB) may arise from the superposition of unresolved astrophysical and cosmological sources \cite{Abbott:1984fp, Allen:1987bk}. 
Unlike other signal propagation, GWs travel through the intervening medium nearly unaltered, thereby bearing the information about its source in a rather preserved manner. This opens up the possibility of exploiting GW signals in extracting information about their origin and the corresponding era during the evolution history of the universe.
The detection of a primordial SGWB would provide a unique and independent probe of the early Universe and its underlying physics, offering access to processes inaccessible to conventional electromagnetic observations. Recent Pulsar Timing Array (PTA) observations have already reported strong evidence for an SGWB in the nano-Hertz frequency band \cite{NANOGrav:2023gor, Antoniadis:2023rey, Reardon:2023gzh, Xu:2023wog}, highlighting the growing observational sensitivity to such backgrounds. In this context, primordial black holes (PBHs), originally proposed in Refs.~\cite{Zeldovich:1967lct, Hawking:1971ei, Carr:1974nx}, provide a compelling framework connecting early-Universe dynamics, dark matter (DM), and GW phenomenology. PBHs are currently studied not only as viable DM candidates \cite{Carr:2020xqk, Green:2020jor, Carr:2021bzv}, but also as potential progenitors of the binary black hole coalescences detected by the LVK collaboration \cite{Bird:2016dcv, Clesse:2016vqa, Sasaki:2016jop}. Formed from the collapse of enhanced small-scale density perturbations upon horizon re-entry during the radiation-dominated era \cite{Hawking:1971ei, Carr:1974nx, Carr:1975qj, Sasaki:2018dmp}, PBHs are directly linked to the primordial power spectrum at scales far smaller than those probed by the Cosmic Microwave Background (CMB) \cite{Josan:2009qn, Green:2020jor}, naturally giving rise to observable signatures such as a stochastic GW background \cite{Matarrese:1997ay, Ananda:2006af, Baumann:2007zm}.

On the other hand, estimating the present-day expansion rate of the universe, quantified by the Hubble parameter $H_0$, remains one of the most promising challenges in modern cosmology. Several conventional probes end up at estimating different values for $H_0$ ranging between  $\sim 67-73~{\rm km\,s^{-1}\,Mpc^{-1}}$, leading to a statistically significant level of tension~\cite{DiValentino:2021izs, Abdalla:2022yfr}
between early-universe measurements, inferred from CMB~\cite{Planck:2018vyc} and Baryon Acoustic Oscillations (BAO)~\cite{eBOSS:2020yzd}, and late-universe direct measurements utilizing the cosmic distance ladder, such as Type Ia supernovae (SNIa) calibrated by Cepheid variables~\cite{Riess:2021jrx}. This persistent discrepancy underscores the critical need for independent, high-precision probes of $H_0$ that do not rely on traditional electromagnetic calibrations~\cite{LIGOScientific:2017adf}.
As it turns out, a distance ladder-independent measurement of the Hubble parameter may serve as a fresh direction to look into the scenerio from a new perspective.
In this work, we explore how correlated GW signals from a common primordial source can enhance the precision of the $H_0$ measurement through GW observations. In particular, we demonstrate that the precision of the $H_0$ measurement can be significantly improved through the observation of distinct GW signals originating from a common primordial source, namely PBHs. To this end, we consider two GW components associated with PBHs: scalar-induced gravitational waves (SIGWs) and the GWs produced by mergers of the same PBH population~\cite{Liu:2021jnw}, thereby providing a correlated multi-band GW signal. The first component, SIGWs, is generated at second order in cosmological perturbation theory due to the nonlinear coupling of first-order scalar curvature perturbations with tensor modes during the radiation-dominated era \cite{Kohri:2018awv, Saito:2008jc, Dom_nech_2020, Domenech:2021ztg}. When such enhanced small-scale curvature perturbations re-enter the Hubble horizon, overdense regions can undergo gravitational collapse to form PBHs if their local density contrast exceeds a critical threshold during the radiation-dominated era \cite{Hawking:1971ei, Carr:1974nx, Carr:1975qj, Sasaki:2018dmp}.  Over cosmic time, these PBHs can form binaries and eventually merge, producing GWs that contribute to the SGWB independently of the induced component. Recent studies have further highlighted the potential role of PBHs in explaining sub-solar mass merger events, such as the recently reported event S251112cm~\cite{Vieira:2026eof}, thereby strengthening the observational motivation for this scenario \cite{Haque:2026yum}.

A key recent insight into the scenario is that GW signals from these correlated primordial sources can appear in two distinct frequency bands with different spectral shapes. This allows the induced GW spectrum to constrain PBH scenarios, and vice versa \cite{Saito:2008jc, Alabidi:2012ex, Nakama:2016enz}. Nevertheless, these PBHs can serve as a possible probe for indirectly measuring the precision of the Hubble parameter $H_0$ by exploiting the connection between these two peak frequencies and the Hubble parameter~\cite{Liu:2021jnw}. Consequently, any future set of missions designed to operate at two distinct frequency bands that remain sensitive to a particular class of PBHs, can act as  a potentially independent probe of the Hubble parameter, when taken in unison.
Our primary intention in this work is to look for such possible combinations of future missions and to investigate their combined prospects as an independent probe of $H_0$ estimating the uncertainties in such measurements. This will in turn give us an idea about the expected precision in the measurement of $H_0$, once the real data from those missions arrive.


By cross-correlating the peak frequencies of the multi-band SGWB signals, specifically, the induced GWs accessible to low-frequency detectors and the binary merger signals observable by high-frequency detectors, we formulate a novel framework to probe the precision of the $H_0$ from future GW observations. As stated earlier, a crucial aspect of this work is the choice of multi-band GW observations that can simultaneously probe these two components. For PBHs in the solar to sub-solar mass range, the induced GW signal typically peaks in the nanohertz frequency band, $\nu \sim 10^{-9}$–$10^{-8}$ Hz \cite{Saito:2008jc}. While current PTA observations such as NANOGrav provide compelling evidence for an SGWB, their sensitivity remains limited for precise estimation of PBH-induced signals over a broad parameter space \cite{NANOGrav:2023hvm, NANOGrav:2023gor}. In contrast, the Square Kilometre Array (SKA), as a next-generation PTA, will significantly enhance timing precision, increase the number of monitored millisecond pulsars, and extend the observation baseline, thereby substantially improving the detection prospects and parameter estimation accuracy for induced GWs in this frequency range \cite{Janssen:2015dca, Weltman:2019neo} (for the sensitivity comparison, see Fig.\ref{fig:Sensitivities}). On the other hand, the merger-induced GW background associated with the same PBH population lies in the frequency range $\nu \sim 1$–$10^3$ Hz \cite{Wang:2016ana, Raidal:2017mfl}. Although current ground-based detectors such as LIGO operate in this band, their sensitivity is not sufficient  to robustly probe the stochastic merger background across the relevant PBH parameter space \cite{KAGRA:2021kbb, KAGRA:2025iso, KAGRA:2025cosmo}. Future GW detectors such as the Einstein Telescope (ET) are expected to achieve orders-of-magnitude improvement in sensitivity in  the same frequency band, along with a significantly wider frequency coverage, (see, Fig. \ref{fig:Sensitivities}), enabling high signal-to-noise detection of both individual merger events and the induced background \cite{Punturo:2010zz, Maggiore:2019uih}. Therefore, the combination of future observations like SKA and ET provides an optimal and complementary observational strategy, targeting the same PBH population across widely separated frequency bands with significantly enhanced detection prospects compared to existing facilities. This complementarity in frequency coverage and sensitivity motivates our choice of SKA and ET as a representative and optimal multi-band observational setup in this work.

As stated earlier, our primary intention with this setup  is to do an estimation of uncertainties in  an  indirect measurement of the Hubble parameter by utilizing  observations at two different frequency bands SKA and ET. This is materialized via the measurement of uncertainties in  the horizon mass at PBH formation and consequently PBH parameters $M_{\rm PBH}$ and $f_{\rm PBH}$ via the correlated GW peak frequencies.
By cross-correlating the peak frequencies of the multi-band SGWB signals, the induced GWs accessible to low-frequency detectors and the binary merger signals observable by high-frequency detectors, we formulate a novel framework to probe the precision of $H_0$ from future GW observations. Importantly, one can express $H_0$ schematically as $H_0 \sim \Omega_{\rm R}^{-1/2}\,\nu_{\rm I}^2/\nu_{\rm M}$, where $\nu_{\rm I}$ and $\nu_{\rm M}$ denote the peak frequencies of the induced and merger GW spectra, respectively, and $\Omega_{\rm R}$ is the present-day radiation energy density fraction. However, this connection is mediated by the PBH parameters $M_{\rm PBH}$ and $f_{\rm PBH}$, implying that any inference of $H_0$ is intrinsically tied to the accuracy with which these parameters can be determined.  To quantify this, we first compute the signal-to-noise ratio to assess the detectability of the GW signals across the PBH parameter space \cite{Allen:1996vm, Allen:1997ad, Flanagan:1997kp}. Building on this, we perform a statistical forecast using the Fisher information matrix formalism to evaluate the precision with which the PBH parameters can be determined \cite{Cutler:1994ys, Poisson:1995ef}. We then propagate these parameter uncertainties through the frequency–horizon-mass relation to estimate the resulting uncertainty in $H_0$. This establishes a direct mapping between GW observables and cosmological precision, explicitly demonstrating how uncertainties in PBH parameters translate into uncertainties in $H_0$. Overall, this multi-band GW framework provides an independent and complementary approach to forecasting the uncertainty in $H_0$. Ultimately, our analysis demonstrates that GWs multi-band observations with SKA and ET can probe a well-motivated region of the PBH parameter space with sufficiently high signal-to-noise ratios  and relatively small parameter uncertainties. As a representative case, we find that the resulting uncertainty in the inferred Hubble parameter, $\delta H_0$, can be constrained at the level $\mathcal{O}(10^{-2} - 1)\ \mathrm{km\,s^{-1}\,Mpc^{-1}}$, for two different approaches, namely, conservative and optimistic, depending on the underlying PBH abundance and mass scale. This highlights that a precision measurement of $H_0$ using the combined capabilities of SKA and ET is not only feasible but also provides a competitive and complementary probe of cosmology.



\section{PBHs as sources for GWs}

\subsection{General aspects of PBHs}

Primordial black holes provide a crucial window to the physics of the early Universe, as they can form directly from the gravitational collapse of enhanced density fluctuations. In particular, during the radiation-dominated (RD) era, overdense regions that re-enter the Hubble horizon may undergo collapse into BHs if their amplitude exceeds a critical threshold, $\delta_{\text{th}}$. This threshold encapsulates the competition between gravitational attraction and pressure gradients in the relativistic plasma. In this regard, the mass of the resulting PBH is set by the horizon scale at the time of collapse. Since causal physics operates within the Hubble volume, the PBH mass is naturally expected to be proportional to the total energy enclosed within this region. Accordingly, the initial PBH mass can be written as
\begin{equation}
M_{\rm in}
=
\gamma \rho_{\rm R}(t_{\rm in})
\frac{4}{3}\pi
\frac{1}{H_{\rm in}^{3}}
=
4\pi\gamma
\frac{M_{\rm P}^2}{H_{\rm in}}
\simeq
1.3~\gamma
\left(
\frac{10^{14}~{\rm GeV}}{H_{\rm in}}
\right) ~{\rm g},
\label{Eq:min}
\end{equation}
where $H_{\rm in}$ and $\rho_{R}(t_{\rm in})$ denote the Hubble expansion rate and the radiation energy density at the time of formation, respectively. The dimensionless parameter $\gamma$ characterizes the collapse efficiency, i.e., the fraction of the horizon mass that ultimately collapses to form a PBH. Although numerical simulations show that $\gamma$ depends sensitively on the perturbation profile and the background cosmology~\cite{Musco:2012au,Musco:2008hv,Hawke:2002rf,Niemeyer:1997mt,Escriva:2021pmf,Escriva:2019nsa,Escriva:2020tak,Escriva:2021aeh}, a commonly adopted benchmark value is $\gamma = (1/3)^{3/2} \simeq 0.2$~\cite{Carr:1974nx}. In this analysis, we adopt this standard value and subsequently demonstrate how our results depend on variations of the $\gamma$ parameter. 
To quantify the abundance of PBHs, we introduce the mass fraction at formation,
\begin{equation}
\beta = \left.\frac{\rho_{\rm PBH}}{\rho_{\rm tot}}\right|_{\rm form}
\simeq 3.7 \times 10^{-9} 
\left(\frac{\gamma}{0.2}\right)^{-1/2} 
\left(\frac{g_{*}(T_{\rm in})}{10.75}\right)^{-1/4} 
\left(\frac{M_{\mathrm{PBH}}}{M_{\odot}}\right)^{1/2} 
f_{\mathrm{PBH}},
\label{eq:beta1}
\end{equation}
where $\rho_{\rm PBH}$ is the energy density of PBHs, $f_{\rm PBH}$ denotes the present fraction of PBHs relative to the total dark matter density, and $g_{*}$ represents the effective number of relativistic degrees of freedom. Assuming PBHs form from the collapse of overdense regions, the mass fraction $\beta$ can be interpreted as the probability that the density contrast exceeds a threshold $\delta_{\rm th}$,
\begin{equation}
\beta = \gamma \int_{\delta_{\rm th}}^{1} P(\delta)\, d\delta.
\end{equation}
For Gaussian fluctuations, this gives
\begin{equation}
\beta(M_{\rm PBH}) \simeq 
\frac{1}{\sqrt{2\pi}}
\frac{\sigma(M_{\rm PBH})}{\delta_{\text{th}}}
\exp\!\left[-\frac{\delta_{\text{th}}^2}{2\sigma^2(M_{\rm PBH})}\right],
\label{eq:beta2}
\end{equation}
which is valid in the limit $\delta_{\text{th}} \gg \sigma(M_{\rm PBH})$. Here, $\sigma({M_{\rm PBH}})$ is the variance of density fluctuations smoothed on the corresponding scale. The variance is related to the primordial curvature power spectrum as
\begin{equation}
\sigma^2(M_{\rm PBH}) = \int d\ln k\, W^2(kR)\, P_\delta(k)
= \int d\ln k\, W^2(kR)\, \frac{16}{81}(kR)^4 \mathcal{P}_{\mathcal{R}}(k),
\end{equation}
where $R$ denotes the comoving radius of the overdense region and $W(kR)$ is the window function relevant for PBH formation. We  use  the relation between the power spectrum of the density contrast, $\mathcal{P}_\delta(k)$, and the primordial curvature perturbation spectrum, $\mathcal{P}_{\mathcal{R}}(k)$, given by~\cite{Wands:2000dp}
\begin{equation}
\mathcal{P}_\delta(k)=\left[\frac{2(1+w)}{5+3w}\right]^2 \,\left(kR\right)^4 \mathcal{P}_\mathcal{R}(k),
\label{eq:curv-matter}
\end{equation}
We note that the PBH abundance is exponentially sensitive to both the threshold $\delta_{\rm th}$ and the amplitude of primordial fluctuations. Consequently, even small changes in these quantities can lead to significant variations in the predicted PBH abundance. We shall assume that the threshold value of the density contrast is given by the following analytical expression~\cite{Harada:2013epa} \footnote{The analytic expression is derived under simplified assumptions of spherical symmetry and relativistic perfect fluid dynamics, and thus does not fully capture non-sphericity and profile dependence; nevertheless, it captures the essential physics of PBH formation and agrees well with numerical simulations, particularly for $w=1/3$~\cite{Escriva:2019nsa}.}
\begin{equation}
\delta_{\rm th}
=\frac{3(1+w)}{5+3\,w}\,\sin^2\!\left(\frac{\pi\,\sqrt{w}}{1+3\,w}\right).
\label{eq:cd}
\end{equation}
Using this expression, one finds that for radiation domination ($w = 1/3$), the threshold reduces to $\delta_{\rm th} \simeq 0.41$, which we will use throughout the analysis.

The abundance and mass distribution of PBHs derived above have direct implications for GW signals. In particular, we focus on two complementary GW sources associated with PBHs. First, PBHs forming binary systems and merging to generate a stochastic GW background from unresolved events. The amplitude and spectral shape of this merger signal depend on the PBH abundance $f_{\rm PBH}$ and mass scale $M_{\rm PBH}$. Second, the large primordial curvature perturbations responsible for PBH formation inevitably induce GWs at second order. These two signals probe complementary physics - merger GWs trace the late-time PBH population, while induced GWs encode the primordial fluctuations at formation. As a result, a combined framework encompassing both the scenarios, taken together, may serve as a powerful probe of the early Universe, which in turn acts a natural reason to revisit the relevant parameters that follow directly from such a combined analysis. This is what we are going to explore at length in the present article.

\subsection{GWs from PBH mergers}

PBHs can form binary systems that eventually merge, producing a GW background. To quantify this signal, we define the present-day GW energy density per logarithmic frequency interval as
\begin{equation}
\Omega_{\rm GW}(\nu) 
= \frac{1}{\rho_c} \frac{{\rm d}\rho_{\rm GW}}{{\rm d}\ln\nu}
= \frac{\nu}{\rho_c} \frac{{\rm d}\rho_{\rm GW}}{{\rm d}\nu},
\end{equation}
where $\nu$ is the observed frequency, $\rho_c = 3H_0^2 M_{\rm P}^2$ is the critical energy density today, with $H_0$ the present-day value of Hubble parameter. The corresponding GW energy density is obtained by integrating the redshifted source-frame spectrum over the merger rate,
\begin{equation}
\Omega_{\mathrm{GW}}^{\rm M}(\nu)
= \frac{\nu}{\rho_c}
\int_0^{z_{\rm sup}} {\rm d}z 
\int {\rm d}M_1 {\rm d}M_2\,
\frac{R(z,M_1,M_2)}{(1+z)H(z)}
\left.\frac{{\rm d}E_{\rm GW}}{{\rm d}\nu_s}\right|_{\nu_s=(1+z)\nu}
\label{eq:BGW}
\end{equation}
where $R(z,M_1,M_2)$ denotes the PBH merger rate per comoving volume and $z_{\rm sup} = \min(z_{\rm max}, \\\nu_c/\nu - 1)$ is determined by the cutoff frequency $\nu_c$. For a binary with component masses $M_1$ and $M_2$, the chirp mass and symmetric mass ratio are defined as
\begin{equation}
M_c = \frac{(M_1 M_2)^{3/5}}{(M_1 + M_2)^{1/5}}, 
\qquad 
\eta = \frac{M_1 M_2}{(M_1 + M_2)^2}.
\end{equation}
In the non-spinning limit, the inspiral--merger--ringdown  GW energy spectrum is approximated by~\cite{Ajith:2007kx, Ajith:2009bn}
\begin{equation}
\frac{{\rm d}E_{\rm GW}}{{\rm d}\nu_s}
= \frac{(\pi G)^{2/3} M_c^{5/3}}{3}
\begin{cases}
\nu_s^{-1/3}, & \nu_s < \nu_1, \\[4pt]
\frac{\nu_s^{2/3}}{\nu_1}, & \nu_1 \le \nu_s < \nu_2, \\[4pt]
\frac{\nu_2^{-4/3}}{\nu_1} 
\frac{\sigma^4 \nu_s^2}{\left(\sigma^2 + 4(\nu_s - \nu_2)^2\right)^2}, 
& \nu_2 \le \nu_s \le \nu_3, \\[4pt]
0, & \nu_s > \nu_3,
\end{cases}
\end{equation}
where the phenomenological frequencies $\nu_1,\nu_2,\sigma,\nu_3$ depend on the total mass $M=M_1+M_2$ and the mass ratio $\eta$. The merger rate is determined by the PBH abundance and mass function, given by~\cite{Raidal:2018bbj,Liu:2018ess,Vaskonen:2019jpv}
\begin{equation}
R(z, M_1, M_2)
= \frac{1.6\times 10^6}{\rm Gpc^{3}\,yr}
f_{\rm PBH}^{\frac{53}{37}}
\left(\frac{t}{t_0}\right)^{-\frac{34}{37}}
\left(\frac{M}{M_\odot}\right)^{-\frac{32}{37}}
\eta^{-\frac{34}{37}}
S(f_{\rm PBH}, M)\,
\psi(M_1)\psi(M_2),
\label{eq:merger_rate}
\end{equation}
 where $t_0$ is the age of the Universe and $S$ accounts for suppression effects due to clustering and Poisson fluctuations. The suppression factor can be approximated as \cite{Hutsi:2020sol}
\begin{equation}
S \simeq 0.24 \left(1 + \frac{2.3\,\sigma_{\rm M}^2}{f_{\rm PBH}^2}\right)^{-21/74},
\end{equation}
where $\sigma_{\rm M}\simeq 0.005$ denotes the variance of matter density perturbations at the time of binary formation. For simplicity, we assume a monochromatic PBH mass function,
\begin{equation}
\psi(M') = \delta(M' - M_{\rm PBH}),
\end{equation}
and restrict to equal-mass binaries,
\begin{equation}
M_1 = M_2 = M_{\rm PBH}.
\end{equation}
In this case, the binary parameters reduce to $M = 2M_{\rm PBH}$, $M_c = 2^{-1/5} M_{\rm PBH}$ and $\eta = 1/4$, and the GW spectrum depends only on a single mass scale $M_{\rm PBH}$. The resulting merger-induced GW spectrum typically exhibits a peak structure. The characteristic peak frequency is set by the PBH mass scale and can be expressed as
\begin{equation}
\nu_{\rm M} = \frac{M_{\rm P}^2}{M_{\rm PBH}} \,\mathcal{C}_{\rm M}(\Theta),
\label{eq:nuB}
\end{equation}
where $\mathcal{C}_{\rm M}(\Theta)$ encodes the dependence on binary parameters. This establishes a direct connection between the merger GW signal and the PBH formation scale.

\subsection{Scalar-induced GWs}
Beyond the GW background from PBH mergers, an additional contribution arises from tensor perturbations generated at second order, sourced by the nonlinear coupling of first-order scalar perturbations. These induced GWs are produced when enhanced curvature perturbations re-enter the horizon and act as a source for tensor modes. In the conformal Newtonian gauge, the perturbed FLRW metric is given by~\cite{Baumann:2009ds}
\begin{equation}
ds^2 = a^2(\tau) \left\{ -(1+2\Phi)d\tau^2 + \left[ (1-2\Psi)\delta_{ij} + \frac{1}{2}h_{ij} \right] dx^i dx^j \right\},
\end{equation}
where $\Phi$ and $\Psi$ are the first-order scalar potentials (Bardeen potentials), and $h_{ij}$ denotes the transverse-traceless tensor perturbations.
\begin{figure}
    \centering
    \includegraphics[scale=0.35]{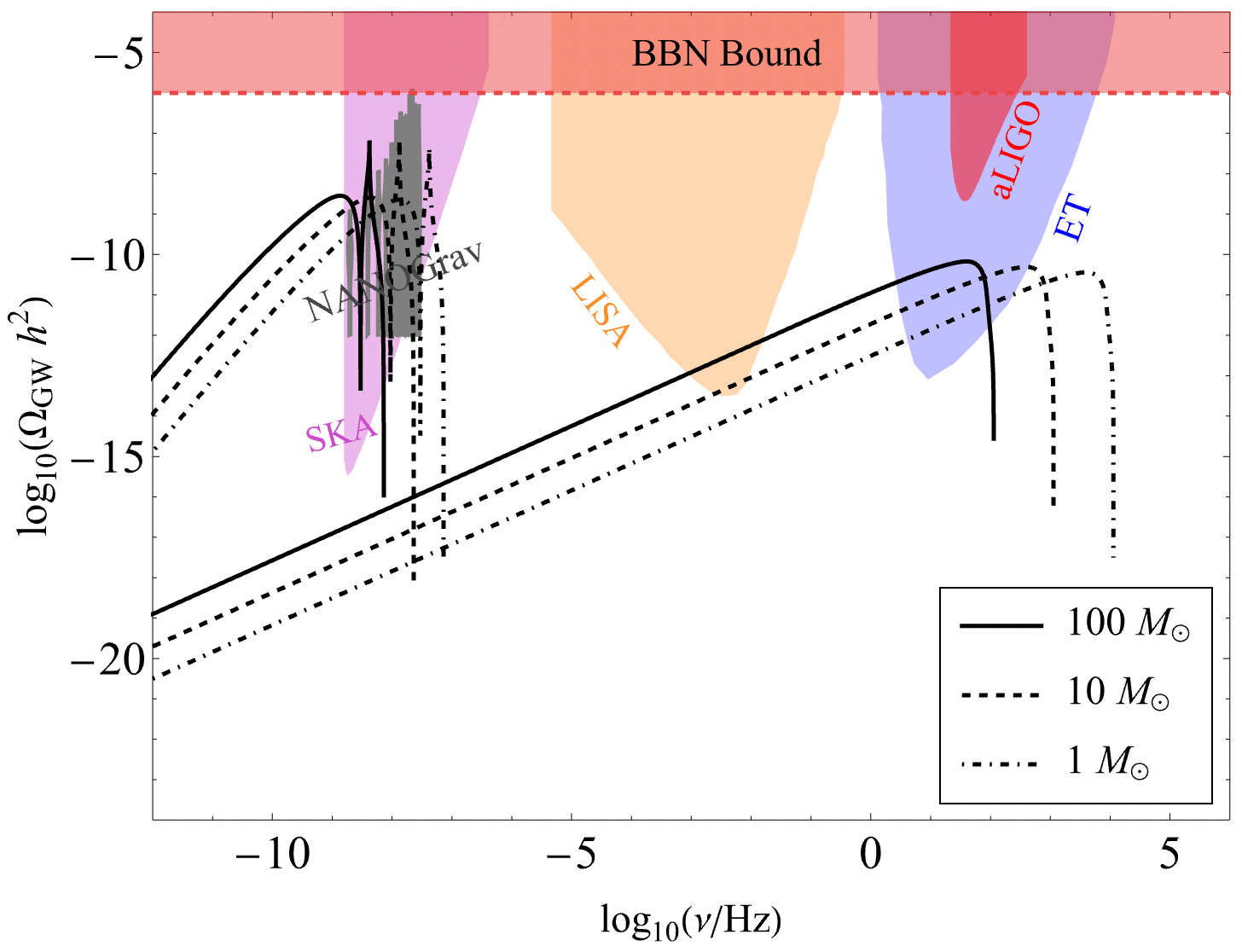}
    \caption{\it{Combined GW spectra, $\Omega_{\rm GW} h^2$, as a function of frequency $\nu$ in Hz. The black curves represent the total GW background from PBHs, including both merger-induced and scalar-induced contributions, for a monochromatic mass distribution with $M_{\rm PBH} = 100\,M_\odot$ (solid), $10\,M_\odot$ (dashed), and $1\,M_\odot$ (dotted), assuming $f_{\rm PBH} = 10^{-3}$. The colored shaded regions indicate the power-law integrated sensitivities of aLIGO, ET, LISA, SKA, and NANOGrav. The horizontal dashed line denotes the BBN upper bound on the GW amplitude arising from constraints on extra relativistic degrees of freedom.}}
    \label{fig:Sensitivities}
\end{figure}
The evolution of the tensor modes is governed by the equation ~\cite{Baumann:2007zm}
\begin{equation}
h_k^{\prime\prime}(\tau) + 2\mathcal{H}h_k^\prime(\tau) + k^2 h_k(\tau) = 4S_k(\tau),
\end{equation}
where primes denote derivatives with respect to conformal time $\tau$, $\mathcal{H}$ is the conformal Hubble parameter, and $S_k(\tau)$ represents the nonlinear source term composed of quadratic combinations of first-order scalar perturbations. Thus the induced GW energy density is given by 
\begin{equation}
\Omega_{\rm GW}^{\rm I} \propto \int_0^\infty {\rm d}v \int_{|1-v|}^{1+v} {\rm d}u \, 
T(u,v)\, \mathcal{P}_{\mathcal{R}}(ku)\, \mathcal{P}_{\mathcal{R}}(kv),
\end{equation}
where $u$ and $v$ are dimensionless momentum ratios, and $T(u,v)$ is the kernel encoding the transfer of scalar perturbations into tensor modes during horizon re-entry. Assuming a sharply peaked dirac-$\delta$ type primordial spectrum~\cite{Kohri:2018awv} \footnote{Although a delta-function spectrum is unphysical and a smooth Gaussian or log-normal profile is more realistic, we adopt this approximation for analytical simplicity. For sufficiently narrow spectra, the delta-function approximation captures the essential features of the induced gravitational wave signal~\cite{Cai:2018dig}.},
\begin{equation}
\mathcal{P}_{\mathcal{R}}(k) = \mathcal{A}_{\mathcal{R}} \, \delta\left(\ln \frac{k}{k_*}\right),
\end{equation}
where $k_*$ denotes the characteristic comoving wavenumber at which the primordial curvature perturbation spectrum is peaked corresponding to the scale responsible for PBH formation and the generation of induced gravitational waves. This simplifies the dimensionless present-day energy density of induced GWs generated during the radiation-dominated era to~\cite{Kohri:2018awv, Dom_nech_2020}
\begin{equation}
\begin{aligned}
\Omega_{\mathrm{GW}}^{\rm I}\,h^2 \simeq \ & 0.43 \left( \frac{g_{*r}}{80} \right)
\left( \frac{g_{*s}}{80} \right)^{-4/3} \Omega_{\rm R}h^2
\frac{3 \mathcal{A}^2_{\mathcal{R}}}{64}
\left( \frac{4 - \tilde{k}^2 }{4} \right)^2 
\tilde{k}^2 \left( 3\tilde{k}^2 - 2 \right)^2 \\
& \times \Bigg[
\pi^2 \left(3\tilde{k}^2 - 2 \right)^2 
\Theta \left( 2 \sqrt{3} - 3 \tilde{k} \right) \\
& \quad + \left( 4 + \left( 3\tilde{k}^2 - 2 \right) 
\ln \left| 1 - \frac{4}{3 \tilde{k}^2} \right| \right)^2
\Bigg]
\Theta \left(2 - \tilde{k} \right),
\end{aligned}
\label{eq:IGW}
\end{equation}
where $\tilde{k} = k/k_*$, $\Theta$ denotes the Heaviside step function, and $g_{*r}$ and $g_{*s}$ are the effective numbers of relativistic degrees of freedom associated with radiation and entropy at the epoch of GW generation, respectively. The amplitude $\mathcal{A}_{\mathcal{R}}$ determines the overall strength of the induced GW signal, while the present radiation energy density is given by $\Omega_{\rm R} h^2 \simeq 4.18 \times 10^{-5}$. Note that the induced GW spectrum exhibits a pronounced peak around $k \sim k_*$, reflecting the underlying curvature perturbation scale. This establishes a direct connection between the primordial perturbations responsible for PBH formation and the resulting GW signal, providing an independent probe of the early Universe. 

Similar to the merger background, the induced GW spectrum exhibits a characteristic peak frequency $\nu_{\rm I}$ set by the horizon scale $k_*$~\cite{Liu:2021jnw},
\begin{equation}
\nu_{\rm I} = k_* \mathcal{C}_{\rm I }(\Theta),
\label{eq:nuI}
\end{equation}
where $\mathcal{C}_{\rm I }(\Theta)$ encodes the dependence on the relevant parameter set $\Theta$. The variance of the density contrast scales as  $\sigma(M_{\rm PBH}) \sim \sqrt{\mathcal{A}_{\mathcal{R}}}$~\cite{Dom_nech_2020}, while the characteristic scale $k_*$ is related to the PBH mass through
\begin{equation}
k_* (M_{\rm PBH}) = 2.35 \times 10^7 \, \mathrm{Mpc}^{-1} \sqrt{\frac{M_\odot}{M_{\rm PBH}}}.
\label{eq:ksM}
\end{equation}
Assuming a monochromatic mass distribution, these relations link the primordial curvature amplitude to the PBH abundance. As a result, the framework is effectively described by two independent parameters - the PBH mass $M_{\rm PBH}$ and its present-day abundance $f_{\mathrm{PBH}}$.

To illustrate the observational prospects of these sources, we present the resulting stochastic GW backgrounds from both coalescing PBH binaries and scalar-induced GWs in Fig.~\ref{fig:Sensitivities}, vis-à-vis the sensitivity curves of current and future GW missions. Since these two signals arise from the same PBH population but peak in widely separated frequency bands, their joint detection requires a multi-band strategy; we therefore focus on SKA and ET, whose enhanced sensitivity relative to current experiments in the frequency ranges of interest makes them ideally suited for this analysis. As evident from Fig.~\ref{fig:Sensitivities}, the benchmark GW spectra naturally intersect the sensitivity curves of these two experiments, making them well-motivated and complementary choices, further supported by their strong projected sensitivity and precision in the near future as discussed in the Introduction. A joint analysis of these two components, based on signal-to-noise ratio estimates and a Fisher matrix approach, will be presented in the next section.

\section{Statistical analysis: Detectability of PBH parameters}\label{sec:forecast}
Before discussing the detectability of the GW signals with future missions, let us briefly outline the methodology adopted to assess their observational prospects. This allows us to keep the analysis self-contained.
\subsection{Signal-to-noise ratio}
\label{subsec:noise_SNR}
The impact of instrumental noise is crucial in determining the detectability of a GW signal. To quantify this, the usual method is to  compute the signal-to-noise ratio, defined as~\cite{Thrane:2013oya,Caprini:2015zlo}
\begin{equation}
\label{eq:snr}
\mathrm{SNR} \equiv \sqrt{\tau_{\rm obs}\;\int_{f_{\rm min}}^{f_{\rm max}} df 
\left(\frac{\Omega_{\rm GW}(f,\{\theta\})h^2}{\Omega_{\rm GW}^{\rm noise}(f)h^2}\right)^2},
\end{equation}
where $\Omega_{\rm GW}(f,\{\theta\})h^2$ denotes the GW energy density spectrum predicted by the model, with $\{\theta\}$ representing the relevant parameters, $\Omega_{\rm GW}^{\rm noise}(f)h^2$ encodes the detector noise spectrum over the frequency range $[f_{\rm min}, f_{\rm max}]$, and $\tau_{\rm obs}$ is the observation time.  As mentioned earlier, we consider future missions SKA and ET,  as representative detectors targeting complementary frequency bands. The corresponding noise spectra are taken from \cite{Schmitz:2020syl}, while the frequency ranges and observation times are summarized in Table~\ref{tab:detector_spec}. For both detectors, we adopt a detection threshold of $\mathrm{SNR} = 1$.
\begin{table}[!ht]
    \centering
    \renewcommand{\arraystretch}{1.2}
    \begin{tabular}{|c|c|c|}
    \hline
    \hline
       \textit{Detector}  & \textit{Frequency range} & $\tau_{\rm obs}$\\
    \hline
       SKA  & $\left[10^{-9}-10^{-7}\right]$ Hz & $20$ years\\
       ET   & $\left[1-10^4\right]$ Hz & $5$ years\\
    \hline
    \hline
    \end{tabular}
    \caption{\it Specification of the detectors under consideration.}
    \label{tab:detector_spec}
\end{table}

\subsection{Fisher forecast analysis}
\label{subsec:fisher}
To quantify the precision with which model parameters can be inferred, we perform a Fisher forecast analysis. While the SNR provides a criterion for detectability, it does not capture the accuracy of parameter estimation. The Fisher matrix formalism, on the other hand, allows us to estimate the expected uncertainties on the model parameters. We begin by discretizing the detector frequency range into logarithmically spaced bins~\cite{Gowling:2021gcy}, such that each bin contributes approximately equally to the total signal. The number of independent modes in the $b$-th bin is given by
\begin{equation}
\label{eq:frequencybin}
n_b \equiv \left[(f_b - f_{b-1})\,\tau_{\rm obs}\right],
\end{equation}
where $b \in [1, N_b]$ and the square brackets denote the integer part. In our analysis, we adopt $N_b = 100$ for all detectors, which ensures the validity of the Gaussian approximation~\cite{Gowling:2021gcy}.

Assuming a Gaussian likelihood for the binned GW signal $\Omega_{\rm sig}(f_b,\{\theta\})$, the likelihood function can be written as~\cite{Dodelson:2003ft}
\begin{equation}
\label{eq:likelihood_fisher}
\mathscr{L}(\theta) = \prod_{b=1}^{N_b} 
\sqrt{\frac{n_b}{2\pi\,\Omega_{\rm n}(f_b)^2}}\,
\exp\left[
-\frac{n_b\left(\Omega_{\rm sig}(f_b,\{\theta\}) - \Omega_{\rm fid}(f_b)\right)^2}
{\Omega_{\rm n}(f_b)^2}
\right],
\end{equation}
where $\Omega_{\rm n}(f_b)$ denotes the noise spectrum and 
$\Omega_{\rm fid}(f_b) \equiv \Omega_{\rm sig}(f_b,\{\theta_{\rm fid}\})$ corresponds to the fiducial model evaluated at parameter values $\{\theta_{\rm fid}\}$. For computational convenience, we work with the logarithm of the likelihood, $\mathcal{L}(\theta) \equiv \ln \mathscr{L}(\theta)$. The Fisher matrix is then defined as~\cite{Dodelson:2003ft}
\begin{equation}
F_{ij} = \left\langle -\,\frac{\partial^2 \mathcal{L}(\theta)}{\partial \theta_i \partial \theta_j} \right\rangle,
\end{equation}
where the expectation value is taken over realizations of the data around the fiducial model. Under the Gaussian approximation, this expression simplifies to
\begin{equation}
F_{ij} = \tau_{\rm obs} \sum_{b=1}^{N_b} 
\frac{2\,\Delta f_b}{\Omega_{\rm n}(f_b)^2}
\frac{\partial \Omega_{\rm sig}(f_b)}{\partial \theta_i}
\frac{\partial \Omega_{\rm sig}(f_b)}{\partial \theta_j},
\end{equation}
where $\Delta f_b = f_b - f_{b-1}$ is the width of the $b$-th frequency bin. The covariance matrix is obtained as the inverse of the Fisher matrix, $C_{ij} = (F_{ij})^{-1}$. The projected $1\sigma$ uncertainties on the model parameters are then given by $\sqrt{C_{ii}}$. In the following, we study the detectability of the PBH parameters $M_{\rm PBH}$ and $f_{\rm PBH}$ based on SNR forecasts and Fisher matrix error estimation, focusing on two future GW missions, SKA and ET. 

\subsection{Detectability of PBH parameters} \label{Pbh parameter estimation}
\begin{figure}
    \centering     
    \includegraphics[scale=0.4]{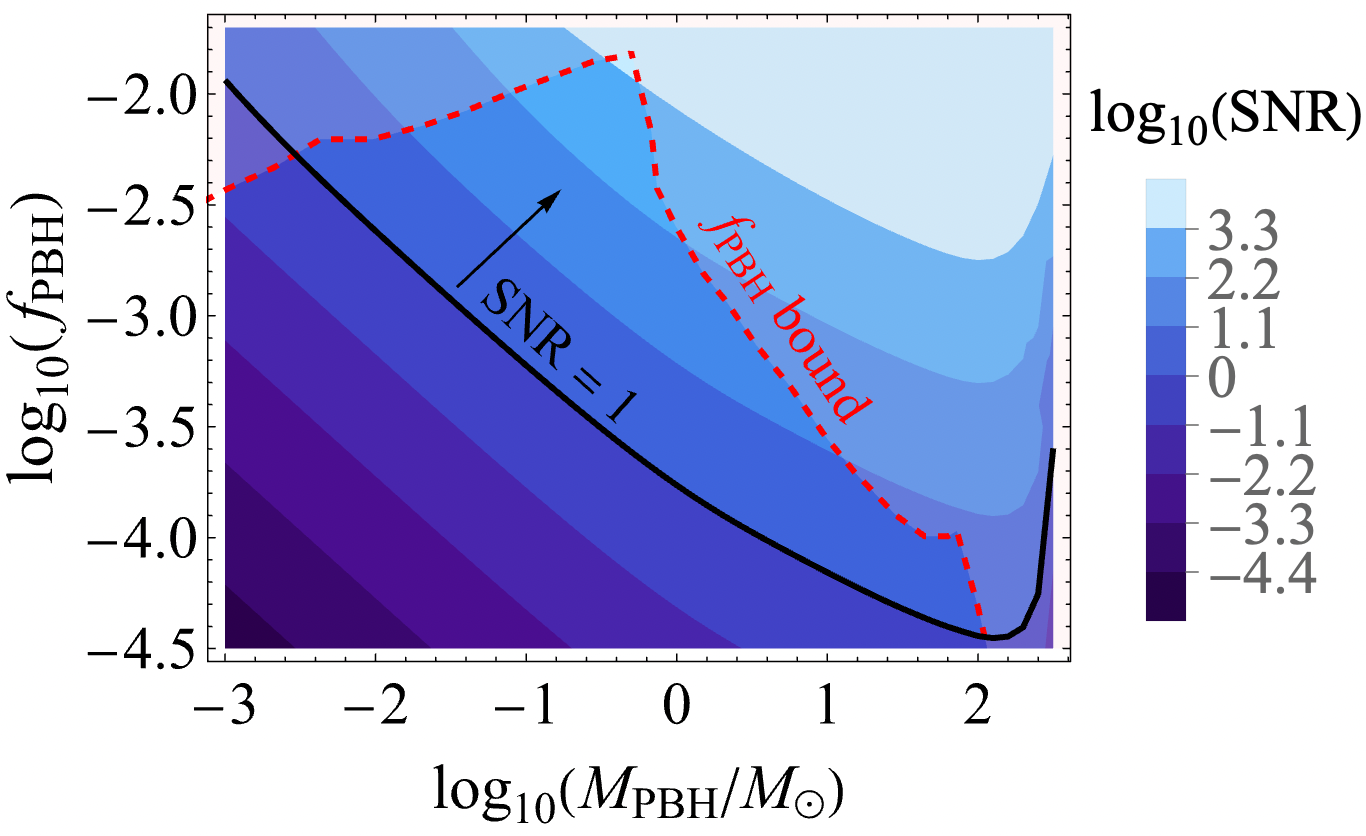}
   \caption{SNR contours in the $(M_{\rm PBH}, f_{\rm PBH})$ plane. The red dashed curve shows current upper limits on $f_{\rm PBH}$, while the black curve denotes the joint $\mathrm{SNR}=1$ threshold obtained by combining SKA and ET observations. Arrows indicate increasing SNR. }
    \label{fig:SNR}
\end{figure}
We determine the detection prospects of PBH parameter space by requiring a signal-to-noise ratio $\mathrm{SNR} \geq 1$, together with the condition that the relative uncertainties on the parameters satisfy
\begin{equation}
\frac{\delta f_{\rm PBH}}{f_{\rm PBH}} \leq 1, 
\qquad 
\frac{\delta M_{\rm PBH}}{M_{\rm PBH}} \leq 1.
\end{equation}
These criteria ensure both detectability and a meaningful estimation of errors in the measurement of the underlying PBH parameters, based on the estimated parameter uncertainties.
Before presenting the forecasts, it is important to place the parameter space in the context of existing observational constraints. Current bounds on $f_{\rm PBH}(M_{\rm PBH})$, assuming a monochromatic mass distribution, arise from a variety of probes~\cite{Carr:2026hot,Oncins:2022ydg,Carr:2016drx,Green:2020jor,Carr:2020xqk}. These include gravitational microlensing~\cite{Niikura:2019kqi,Niikura:2017zjd,Mroz:2024mse}, extragalactic gamma-ray backgrounds~\cite{Carr:2026hot}, and gravitational-wave observations~\cite{Nitz:2021vqh,Kavanagh:2018ggo}. Microlensing constraints from OGLE impose strong limits in the sub-solar mass range, assuming the observed events toward the LMC originate from stellar populations in the LMC or the Milky Way disk~\cite{Mroz:2024mse}. These constraints serve as a reference for interpreting the projected sensitivities from SKA and ET.

In our analysis, SKA primarily probes the induced GW signal at low frequencies, while ET is sensitive to the merger-induced background at higher frequencies. The combination of these two observations  allows us to explore complementary regions of the PBH parameter space, leading to detection aspects of both $M_{\rm PBH}$ and $f_{\rm PBH}$. We emphasize that the condition $\mathrm{SNR} \geq 1$ corresponds to a joint detection, i.e., the parameter region satisfies $\mathrm{SNR} \geq 1$ when combining the contributions from both SKA and ET. The same procedure is followed in the Fisher analysis, where the relative uncertainties, such as $\delta f_{\rm PBH}/f_{\rm PBH}$ and $\delta M_{\rm PBH}/M_{\rm PBH}$, are computed by combining the full GW spectrum, including both induced and merger contributions, along with the instrumental noise from SKA and ET.

\begin{figure}
    \centering
    \begin{minipage}{0.495\textwidth}
        \includegraphics[width=\linewidth]{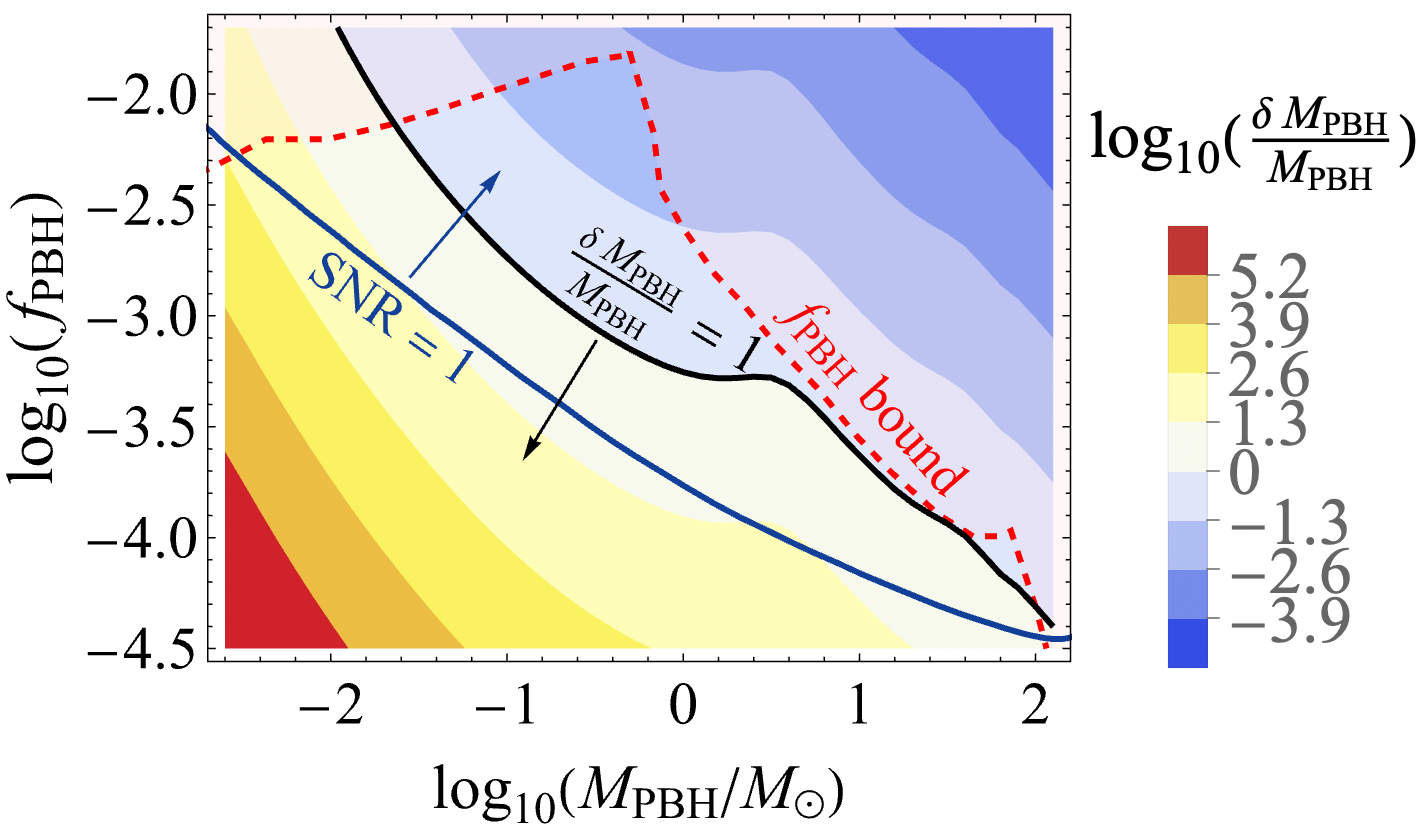}
    \end{minipage}
    \hfill
    \begin{minipage}{0.495\textwidth}
        \includegraphics[width=\linewidth]{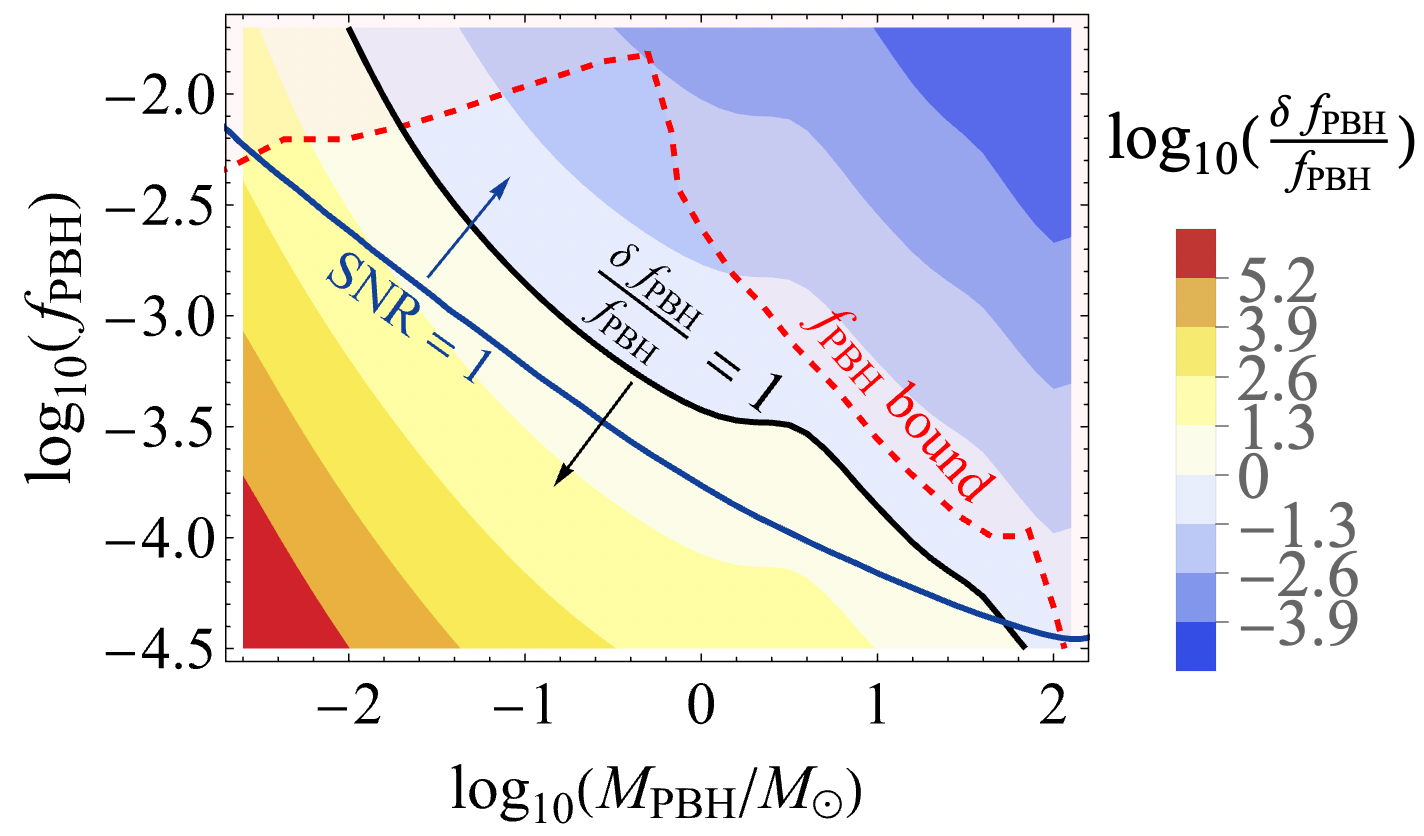}
    \end{minipage}
    \caption{Relative uncertainties on the PBH parameters obtained from Fisher analysis. The blue solid line indicates the $\mathrm{SNR}=1$ detection threshold, with arrows marking the detectable region. The black solid contours correspond to $\delta \theta_i/\theta_i = 1$ (e.g., $\delta M_{\rm PBH}/M_{\rm PBH} = 1$ or $\delta f_{\rm PBH}/f_{\rm PBH} = 1$), while the black arrows indicate the direction of increasing relative uncertainty. The enclosed regions identify the parameter space where the PBH parameters can be constrained with meaningful precision from projected GW observations.}\label{fig:Fisher}
\end{figure}

The parameter space satisfying $\mathrm{SNR} \geq 1$ is shown in Fig.~\ref{fig:SNR}, where the black arrow indicates the region of higher SNR. We find that the allowed region corresponds to $M_{\rm PBH} \in (0.003,\,100)\,M_\odot$ and $f_{\rm PBH} \in (10^{-2.4},\,10^{-4.5})$. On the other hand, imposing the condition on the relative uncertainties, $\delta f_{\rm PBH}/f_{\rm PBH} \leq 1$ and $\delta M_{\rm PBH}/M_{\rm PBH} \leq 1$, leads to a more restricted parameter space (see Fig.~\ref{fig:Fisher}). In this case, the region satisfying both conditions is given by $M_{\rm PBH} \in (0.02,\,100)\,M_\odot$ and $f_{\rm PBH} \in (10^{-2.4},\,10^{-4})$.

Having identified the PBH parameter space with higher detectability from both GW signals using the aforementioned missions, we next connect the peak frequencies of the induced and merger GW spectra to the $H_0$. The estimation of the corresponding uncertainty on $H_0$, obtained by propagating the Fisher-derived parameter uncertainties while considering only instrumental noise, will be presented in the following section.

\section{Estimation of uncertainties in $\mathbf{H_0}$ measurements}

\subsection{$H_0$ from peak frequencies}

As stated earlier, to determine the uncertainty in $H_0$ through multi-band GWs, we need to relate the peak frequency of the GW spectrum to $H_0$. This requires connecting $H_0$ to the horizon scale at the time of PBH formation and the associated comoving wavenumber. Since PBHs originate from the collapse of Hubble-sized regions, their mass is set by the horizon scale at that epoch. Accordingly, the horizon mass can be expressed as $M_{\rm H} = \frac{4}{3}\,\pi \rho H^{-3}$. Moreover, during the radiation-dominated era, the energy density scales as $\rho_{\rm R} \propto g_{*r}(T)\, T^4$. Imposing entropy conservation, $g_{*s}(T)\, T^3 a^3 = \mathrm{constant}$, the scale factor can be related to its present-day value as
\begin{equation}
\frac{a}{a_{\rm 0}} = \left( \frac{g_{*s}(T_0)}{g_{*s}(T)} \right)^{1/3} \frac{T_0}{T}.
\end{equation}
The corresponding Hubble parameter is given by
\begin{equation}
H = \sqrt{\frac{\pi^2}{90}} \, \frac{\sqrt{g_{*r}(T)}\, T^2}{M_{\rm P}},
\label{eq:a_H_evol}
\end{equation}
so that the comoving wavenumber, defined as $k \equiv aH$, can be written as \footnote{Here, we  set the present day scale factor $a_{\rm 0}=1$.}
\begin{equation}
k = \left( \frac{g_{*s}(T_0)}{g_{*s}(T)} \right)^{1/3} 
\frac{T_0}{M_{\rm P}} 
\left( \frac{\pi^2}{90} g_{*r}(T) \right)^{1/2} T.
\label{eq:k_comoving}
\end{equation}
Using Eqs.~\eqref{eq:a_H_evol} and \eqref{eq:k_comoving}, the horizon mass can then be expressed in terms of the comoving wavenumber $k$ as~\cite{Liu:2021jnw}
\begin{equation}
M_H(k) = \frac{1}{2} \left(\frac{k}{M_{\rm P}}\right)^{-2} 
\left( \frac{g_{*s}(T_0)}{g_{*s}(T)} \right)^{2/3} 
\left( \frac{g_{*r}(T)}{g_{*r}(T_0)} \right)^{1/2} 
H_0 \sqrt{\Omega_{\rm R}},
\label{eq:MH_k}
\end{equation}
Turning to the spectral properties of the induced and merger GW signals, their peak frequencies, $\nu_{\rm I}$ (Eq.~\ref{eq:nuI}) and $\nu_{\rm M}$ (Eq.~\ref{eq:nuB}), can be related as \footnote{The parameter $\Theta$ characterizes the shape of the primordial curvature power spectrum is effectively determined by the PBH parameters, i.e., $\Theta \equiv \Theta(M_{\rm PBH}, f_{\rm PBH})$.}
\begin{equation}
\frac{\nu_{\rm I}^2}{\nu_{\rm M}} = \left(\frac{k_{*}}{M_{\rm P}}\right)^2 M_{*} 
\frac{[\mathcal{C}_{\rm I}(M_{\rm PBH}, f_{\rm PBH})]^2}{\mathcal{C}_{\rm M}(M_{\rm PBH}, f_{\rm PBH})},
\label{eq:peak_freq_ratio}
\end{equation}
where $M_{*} \equiv M_H(k_{*}) = \gamma M_{\rm PBH}$, with $\gamma$ denoting the collapse efficiency as mentioned earlier. Substituting Eq.~\eqref{eq:MH_k}, the above relation simplifies to~\cite{Liu:2021jnw}
\begin{align}
\frac{\nu_{\rm I}^2}{\nu_{\rm M}} &= H_0 \sqrt{\Omega_{\rm R}} 
\left[ \frac{1}{2} 
\left( \frac{g_{*s}(T_0)}{g_{*s}(T_{*})} \right)^{2/3} 
\left( \frac{g_{*r}(T_{*})}{g_{*r}(T_0)} \right)^{1/2} 
\right] 
\frac{[\mathcal{C}_{\rm I}(M_{\rm PBH}, f_{\rm PBH})]^2}{\mathcal{C}_{\rm M}(M_{\rm PBH}, f_{\rm PBH})} \nonumber \\
&\equiv H_0 \sqrt{\Omega_{\rm R}} \, Y(M_{\rm PBH}, f_{\rm PBH})\,,
\label{eq:master_H0_relation}
\end{align}
where $Y(M_{\rm PBH}, f_{\rm PBH})$ encodes the ratio of the effective relativistic degrees of freedom and the spectral shape of the GW signal, as defined in the above equation. In the next section, we estimate the uncertainty in the $H_0$, which is linked to the uncertainties in the peak frequencies of the GW spectra. This is achieved by propagating the Fisher forecast uncertainties of the PBH parameters, described in Sec.~\ref{Pbh parameter estimation}.


\subsection{Forecasts on the uncertainties in $H_0$ measurements}
\begin{figure}
    \centering
    \begin{minipage}{0.48\textwidth}
        \centering
        \includegraphics[height=4.9cm]{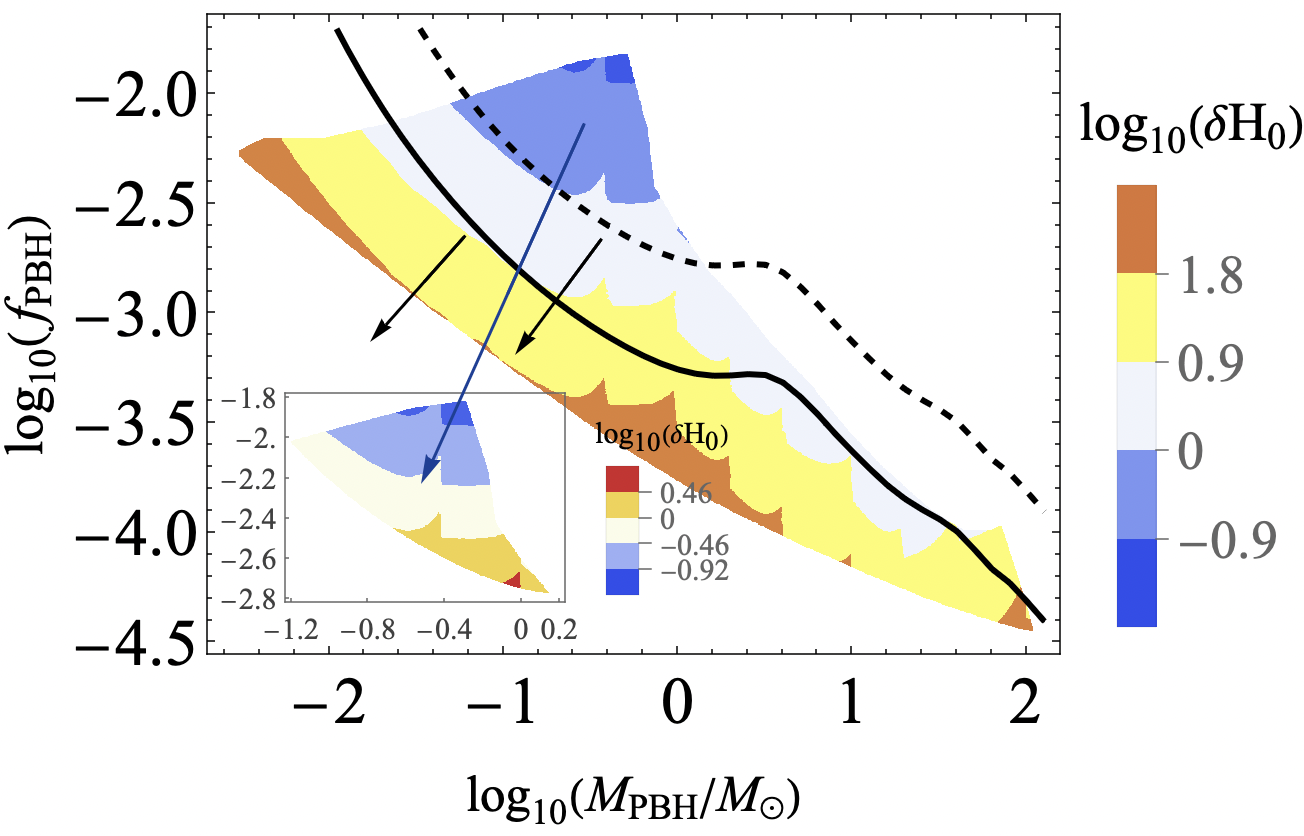}
    \end{minipage}
    \hfill
    \begin{minipage}{0.48\textwidth}
        \centering
        \includegraphics[height=4.85cm]{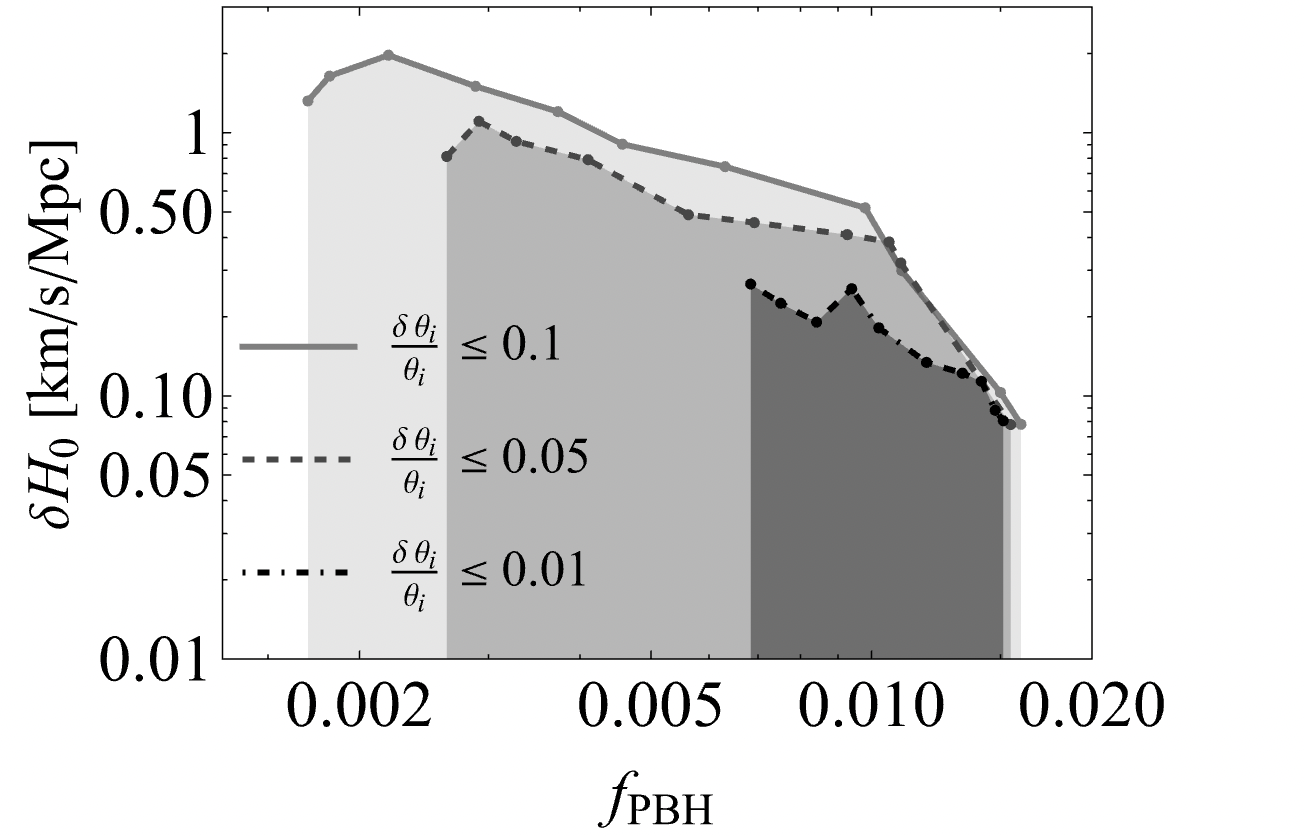}
    \end{minipage}
    \caption{\it Left panel: Absolute uncertainty on the $H_0$, $\delta H_0$, in the PBH parameter space $(M_{\rm PBH}, f_{\rm PBH})$ for regions satisfying $\mathrm{SNR} > 1$, derived from the joint multi-band observation of PBH merger and induced GW signals. A fiducial value $H_0 = 67.66~{\rm km\,s^{-1}\,Mpc^{-1}}$ is assumed.The black solid (dashed) curves represent the contours with $\delta \theta_i/\theta_i = 1~(0.1)$. Right panel: $\delta H_0$ as a function of $f_{\rm PBH}$, illustrating the achievable precision for different relative error on the PBH parameters, with solid, dashed and dot-dashed curves corresponding to $\delta \theta_i/\theta_i \leq ( 0.1,\,0.05,\,0.01)$, respectively.}
    \label{fig:dH0contour}
\end{figure}

Current measurements of the $H_0$ exhibit a significant tension between early and late universe probes~\cite{Planck:2018vyg,Riess:2021jrx}. This highlights the importance of developing independent and complementary approaches to constrain $H_0$, particularly those that rely on entirely different physical observables. In this context, a multi-band GW framework offers a promising avenue, as it enables cross-validation of cosmological parameters across different frequency ranges and observational channels once data from future detectors become available.

In light of this consideration, we investigate multi-band GW observations as an independent probe of the uncertainty in $H_0$, focusing on assessing the achievable precision by exploring a broad range of fiducial values consistent with current observational bounds. To this end, we exploit the relation between the GW peak frequencies and the horizon mass to quantify the corresponding uncertainty in $H_0$. Based on the peak-frequency relation given by Eq.~(\ref{eq:master_H0_relation}),  an indirect estimate of $H_0$ could be obtained as 
\begin{equation}
H_0 = \Omega_{\rm R}^{-1/2} \frac{\nu_{\rm I}^2}{\nu_{\rm M}} \, Y(M_{\rm PBH}, f_{\rm PBH})^{-1}.
\label{eq:H0_estimator}
\end{equation}
Applying standard error propagation and combining the fractional uncertainties in quadrature, the corresponding precision is obtained from Eq.~\eqref{eq:H0_estimator}, leading to
\begin{equation}
\left( \frac{\delta H_0}{H_0} \right)^2 \simeq 
\left( \frac{\delta \nu_{\rm I}}{\nu_{\rm I}} \right)^2 +
\left( \frac{\delta \nu_{\rm M}}{\nu_{\rm M}} \right)^2 +
\left( \frac{\delta Y}{Y} \right)^2 +
\left( \frac{\delta \Omega_{\rm R}}{\Omega_{\rm R}} \right)^2\,,
\label{eq:H0_error}
\end{equation}
where the contribution from $\Omega_{\rm R}$ is negligible. The dominant contribution to the uncertainty in $H_0$ arises from the precision with which the peak frequencies $\nu_{\rm I}$ and $\nu_{\rm M}$ and the spectral factor $Y$ can be determined. All these quantities depend on the PBH parameters $M_{\rm PBH}$ and $f_{\rm PBH}$. Consequently, the relative error  $\delta \nu_{\rm I}/\nu_{I}$, $\delta \nu_{\rm M}/\nu_{\rm M}$, and $\delta Y/Y$ can be expressed in terms of the PBH parameters, allowing Eq.~\eqref{eq:H0_error} to be recast in terms of $\delta \theta_i / \theta_i$, with $\theta_i \in \{M_{\rm PBH},\, f_{\rm PBH}\}$ obtained from the Fisher analysis in Sec.~\ref{Pbh parameter estimation}.

Since we have already estimated the relative uncertainties in the PBH parameters, it is straightforward to determine how these propagate into the uncertainty of the $H_0$. To this end, first we assume, as a representative example, a fiducial value $H_0 = 67.66~{\rm km\,s^{-1}\,Mpc^{-1}}$, consistent with the Planck 2018 results~\cite{Planck:2018vyg}. The resulting uncertainty on the $\delta H_0$, is shown in the parameter-space contour plot in Fig.~\ref{fig:dH0contour}. However, as already stated, the above fiducial value has been taken just as a demonstrative example. The fiducial dependence of error estimation, by taking different fiducial values for $H_0$ within its observational bounds, have been performed and discussed at length in the next section.

An important point to note is that, even within the region satisfying $\mathrm{SNR} \geq 1$ and $\delta \theta_i / \theta_i \leq 1$, the corresponding uncertainty in $H_0$ may not be sufficiently precise in certain regions of the PBH parameter space, characterized by $M_{\rm PBH}$ and $f_{\rm PBH}$. This motivates us to further investigate the conservative approach to PBH parameter estimation, focusing on regions where relatively tighter constraints can be achieved, namely $\delta \theta_i / \theta_i \leq 0.1$, in order to obtain physically meaningful estimates of $H_0$, as illustrated in Fig.~\ref{fig:dH0contour}. Now tightening the relative uncertainty from $\delta \theta_i / \theta_i = 0.1$ to $0.01$ leads to a reduction in $\delta H_0$ from $\lesssim 2$ to $\lesssim 0.2~{\rm km\,s^{-1}\,Mpc^{-1}}$ in optimistic scenarios. However, this uncertainty depends sensitively on both $M_{\rm PBH}$ and $f_{\rm PBH}$. In particular, increasing $f_{\rm PBH}$ results in an $\mathcal{O}(10)$ reduction in the uncertainty. For example, fixing $\delta \theta_i / \theta_i = 0.1$, increasing $f_{\rm PBH}$ from $10^{-3}$ to $10^{-2}$ reduces the Hubble parameter uncertainty from $\delta H_0 \sim 1$ to $\sim 0.1~{\rm km\,s^{-1}\,Mpc^{-1}}$ (see, for instance, right panel of Fig.\ref{fig:dH0contour}). To provide a concise summary of the PBH parameter space of interest and the corresponding uncertainty in the $H_0$, we tabulate the results in Table~\ref{tab:H0_results}.
\begin{table}[t]
\centering
\small
\setlength{\tabcolsep}{6pt}
\renewcommand{\arraystretch}{1.3}
\begin{tabular}{c c c c}
\hline
$\delta \theta_i/\theta_i$ & $M_{\rm PBH}\,[M_\odot]$ & $f_{\rm PBH}$ & $\delta H_0\,[{\rm km\,s^{-1}\,Mpc^{-1}}]$ \\
\hline
$\le 0.1$  & $0.06\!-\!1.4$ & $1.7\!\times\!10^{-3}\!-\!1.6\!\times\!10^{-2}$ & $\lesssim 2$ \\
$\le 0.05$ & $0.08\!-\!1.1$ & $2.6\!\times\!10^{-3}\!-\!1.5\!\times\!10^{-2}$ & $\lesssim 1$ \\
$\le 0.01$ & $0.2\!-\!0.7$  & $6.8\!\times\!10^{-3}\!-\!1.5\!\times\!10^{-2}$ & $\lesssim 0.25$ \\
\hline
\end{tabular}
\caption{\it Absolute uncertainty on the $\delta H_0$, derived from multi-band GW observations for representative PBH parameter ranges.}
\label{tab:H0_results}
\end{table}
\begin{figure}
    \centering
    \includegraphics[scale=0.35]{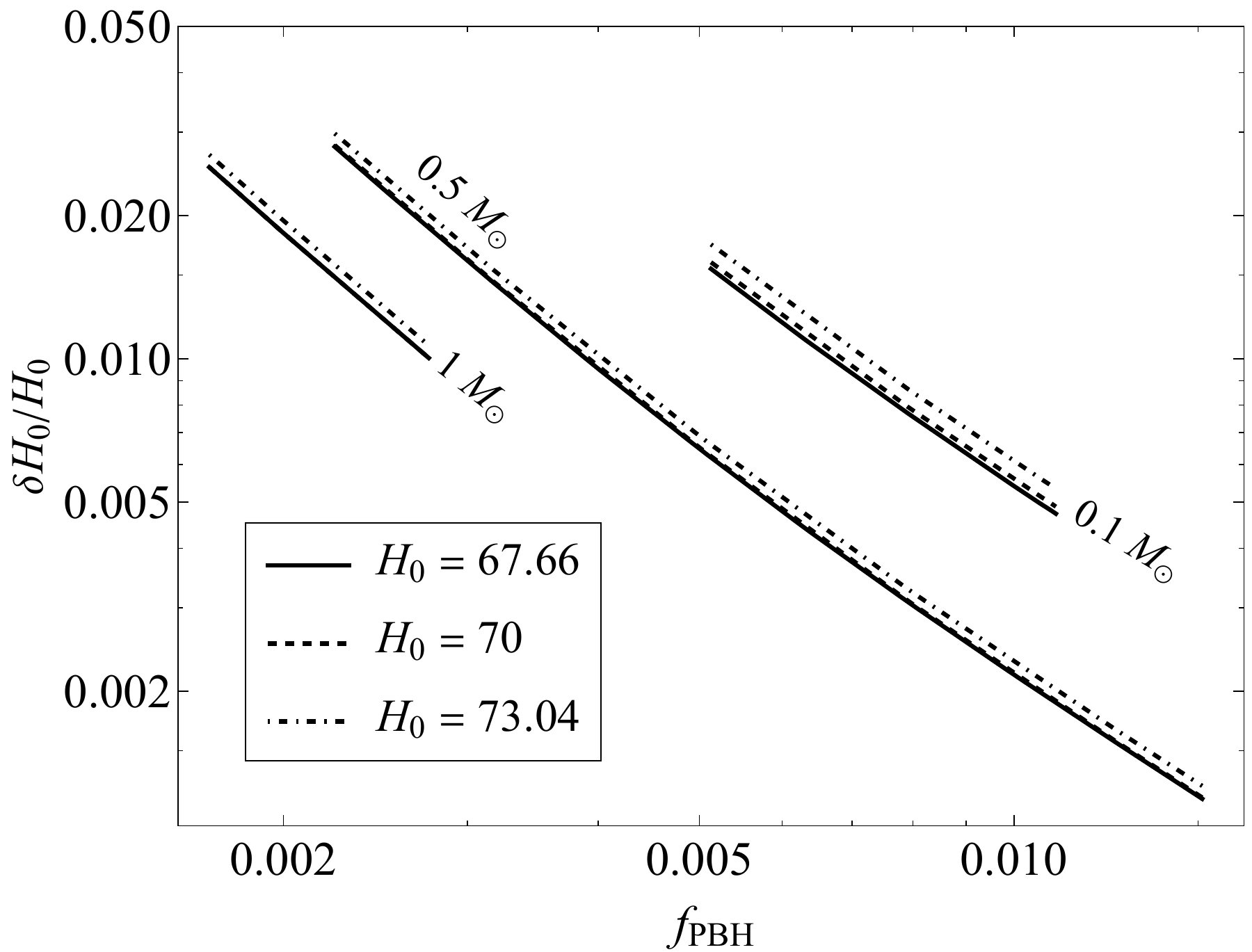}
    \caption{Relative uncertainty on the $H_0$, $\delta H_0/H_0$, as a function of $f_{\rm PBH}$ for three fiducial values $H_0 = 67.66$, $70$, and $73.04~{\rm km\,s^{-1}\,Mpc^{-1}}$. Results are shown for representative PBH masses $M_{\rm PBH} = 1$, $0.5$, and $0.1~M_\odot$. The near overlap of the curves indicates that the fractional precision is largely independent of the fiducial choice of $H_0$.}
    \label{fig:H0_fiducials}
\end{figure}

In summary, this formulation highlights the role of detector sensitivity in setting observational thresholds. In particular, for regions of the PBH parameter space where the relative uncertainties satisfy $\delta \theta_i / \theta_i \leq 0.01$, the corresponding uncertainty in the $H_0$ can be reduced to $\delta H_0 \lesssim 0.1~{\rm km\,s^{-1}\,Mpc^{-1}}$, making the scenario especially promising. Such a level of precision would enable an independent test of the current discrepancies in the measured values of $H_0$~\cite{DiValentino:2021izs,Riess:2021jrx} between early and late universe measurements, providing a complementary cosmological probe that is independent of the cosmic distance ladder~\cite{H0DN:2025lyy}.
As already mentioned, the above estimation of error have been done based on a particular fiducial choice of $H_0$. In the following, we examine how the uncertainty $\delta H_0$ is further affected by variations in the fiducial value of the $H_0$, keeping within its observational bounds, and also of the collapse efficiency, which are key input parameters in our analysis.

\section{Sensitivity of the uncertainties in $\mathbf{H_0}$  to model assumptions}

Let us begin by examining the sensitivity of the present analysis to the choice of the fiducial values of $H_0$. To assess the robustness of our results, we evaluate the predicted relative uncertainty, $\delta H_0 / H_0$, for three representative values spanning the current observational bounds ~\cite{Abdalla:2022yfr,DiValentino:2021izs}, namely $H_0 = 67.66$~\cite{Planck:2018vyg}, $70$, and $73.04~{\rm km\,s^{-1}\,Mpc^{-1}}$~\cite{Riess:2021jrx}. The corresponding results have been presented in Fig.~\ref{fig:H0_fiducials}. As can be found from Fig.~\ref{fig:H0_fiducials}, the resulting relative uncertainties are nearly identical for all three choices, exhibiting a strong overlap across the parameter space. This indicates that our results are largely insensitive to the fiducial value of $H_0$ within the currently allowed range, and the inferred uncertainty remains effectively unchanged over the range of values of the Hubble parameter as long as it lies within the current observational bounds.
\begin{figure}
    \centering
    \begin{minipage}{0.495\textwidth}
        \includegraphics[width=\linewidth]{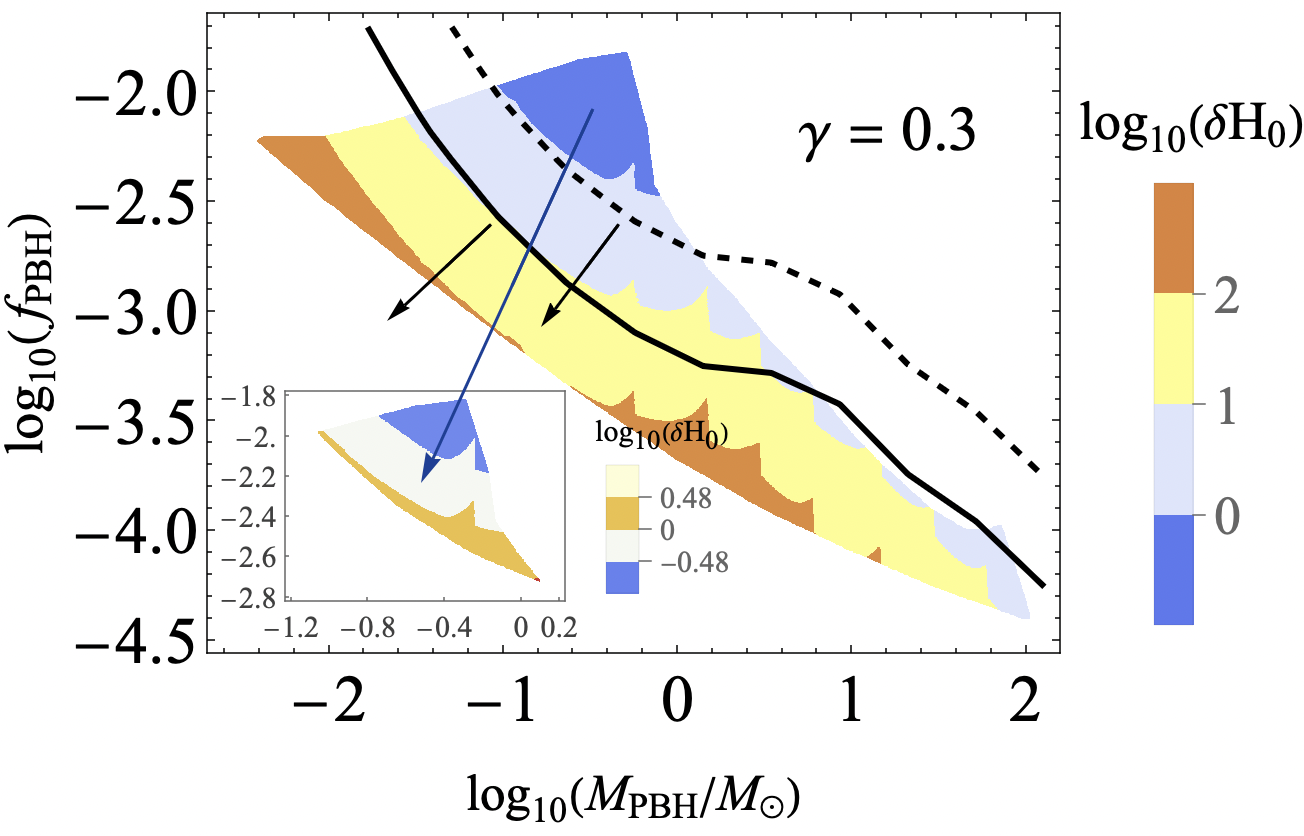}
    \end{minipage}
    \hfill
    \begin{minipage}{0.495\textwidth}
        \includegraphics[width=\linewidth]{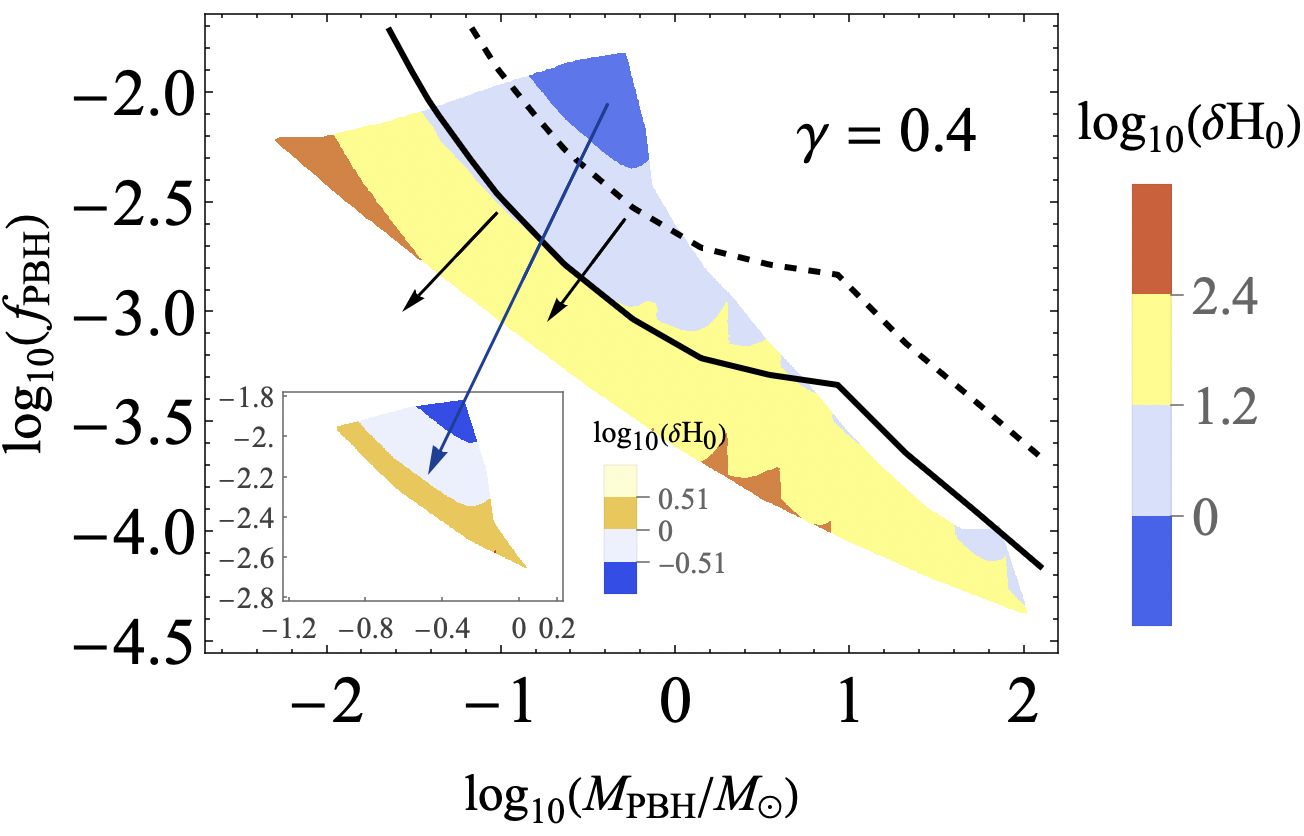}
    \end{minipage}
    \caption{Absolute uncertainty on $\delta H_0$, for two PBH collapse efficiencies. The left (right) panel corresponds to $\gamma = 0.3~(0.4)$, illustrating the impact of collapse efficiency on the inferred precision. The black solid (dashed) curves represent the contours with $\delta \theta_i/\theta_i = 1~(0.1)$.}\label{fig:dH0_gammap3p4}
\end{figure}

\begin{figure}
    \centering
    \includegraphics[scale=0.35]{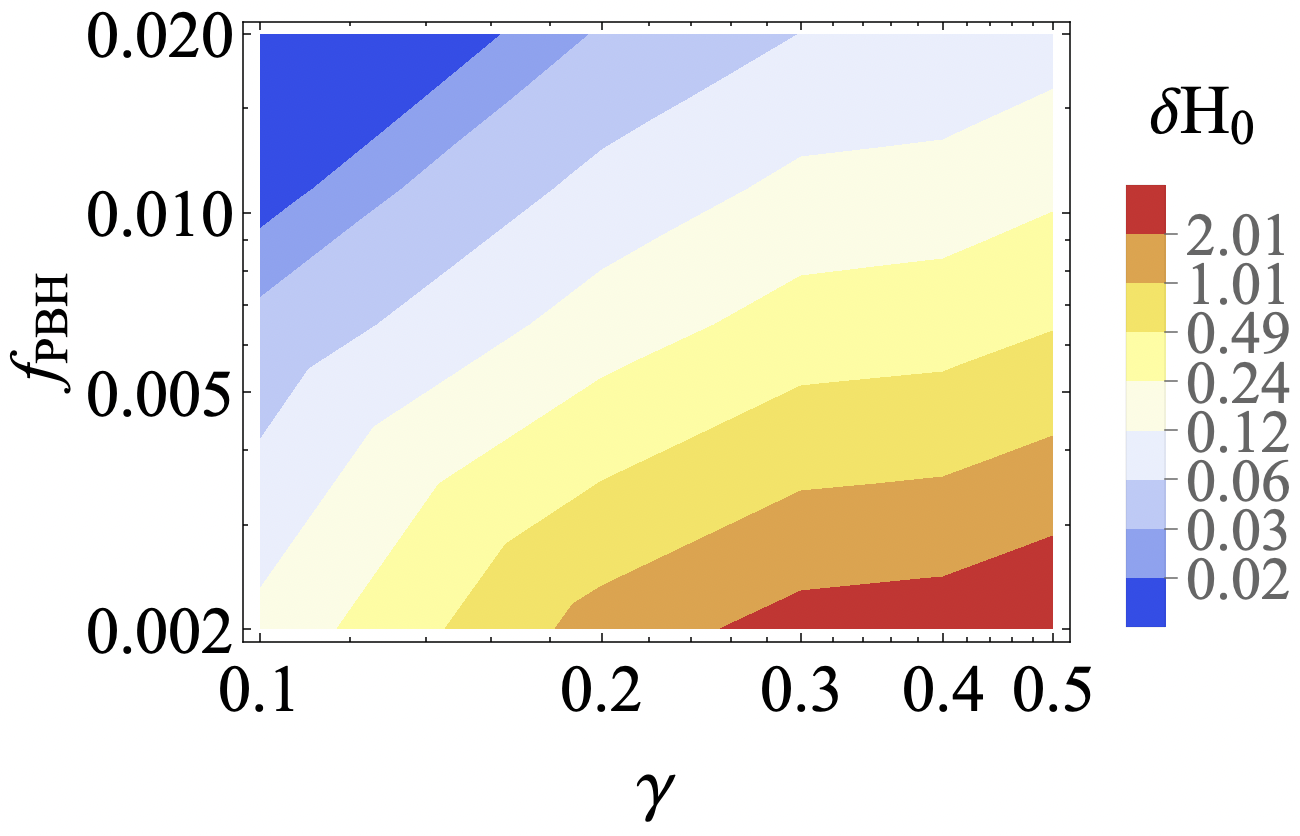}
    \caption{The absolute uncertainty on $H_0$, $\delta H_0$, shown in the $(f_{\rm PBH}, \gamma)$ plane. The PBH mass is fixed to $M_{\rm PBH} = 1~M_\odot$, and a fiducial value $H_0 = 67.66~{\rm km\,s^{-1}\,Mpc^{-1}}$ is assumed.}
    \label{fig:dH0_gamma}
\end{figure}

Further, so far our parameter forecasts have assumed a standard PBH collapse efficiency $\gamma \simeq 0.2$~\cite{Carr:2009jm, Sasaki:2018dmp}, commonly adopted for PBH formation during a radiation-dominated epoch. This choice provides a useful baseline corresponding to simple spherical collapse. However, the exact value of $\gamma$ depends on the detailed collapse dynamics of density fluctuations. In particular, it can be affected by the equation of state at horizon reentry (e.g., near the QCD phase transition), the profile of the primordial curvature perturbations, and deviations from spherical symmetry. Recent numerical studies~\cite{Musco:2008hv,Musco:2012au,Hawke:2002rf} indicate that when the perturbation amplitudes are close to the threshold, i.e., $\delta - \delta_{\rm th} < 10^{-2}$, the collapse efficiency $\gamma$ varies roughly between $0.1$ and $1$. This motivates us to do an investigation of the sensitivity of our results to the choice of $\gamma$. For illustration, we vary it within the range $0.1$–$0.5$ to assess the uncertainty in $H_0$.  In Fig.~\ref{fig:dH0_gammap3p4}, we show the absolute uncertainty on $H_0$, for representative values $\gamma = 0.3$ and $0.4$. Since the collapse efficiency determines the relation between the PBH mass and the horizon mass, $M_{\rm PBH} \simeq \gamma M_H$, varying $\gamma$ modifies the mapping between the observable GW peak frequencies ($\nu_{\rm I}$ and $\nu_{\rm M}$) and the PBH parameters $M_{\rm PBH}$ and $f_{\rm PBH}$. Increasing $\gamma$ slightly reduces the region of parameter space satisfying $\mathrm{SNR} > 1$ and $\delta \theta_i / \theta_i \leq 0.1$, although this shift is relatively mild. Nevertheless, even this modest change leads to a noticeable variation in the inferred uncertainty $\delta H_0$, as illustrated in Fig.~\ref{fig:dH0_gamma}. For instance, fixing $M_{\rm PBH} = 1\,M_{\odot}$ and a given value of $f_{\rm PBH}$, increasing $\gamma$ shifts $\delta H_0$ toward larger values. More generally, varying $\gamma$ over the range $0.1$--$0.5$ induces an $\mathcal{O}(10)$ change in the $H_0$ uncertainty.

Overall, our analysis shows that the results are largely insensitive to the choice of the fiducial values of $H_0$, while exhibiting a moderate dependence on the collapse efficiency $\gamma$. This underscores the importance of accurately modeling PBH formation dynamics when interpreting GW-based constraints on the uncertainty of $H_0$.

\section{Summary and Outlook}

In this article, we estimated the uncertainties in the measurement of the Hubble parameter $H_0$ using multi-band gravitational wave observations sourced by PBHs. In particular, we focussed on two future missions, SKA and ET, which are sensitive to distinct frequency bands for GWs, to act as possible detectors in a combined framework. The key idea relies on the fact that two complementary GW signals, namely, PBH merger GWs (probed by ET) AND induced GWs (probed by SKA), carry independent information that can be combined to extract the Hubble parameter in a self-consistent manner.

By performing a joint analysis of these two GW signals originated from a common source, we investigated the detectability of the PBH parameters, $M_{\rm PBH}$ and $f_{\rm PBH}$, based on combined forecasts on the detectors SKA and ET. As shown in Fig.~\ref{fig:Sensitivities} and Fig.~\ref{fig:SNR}, combining low-frequency observations from SKA with high-frequency measurements from ET, we estimated the  PBH parameter space  accessible to both of them. Imposing the joint detection criterion $\mathrm{SNR} \geq 1$ leads to a well-defined region in the $(M_{\rm PBH}, f_{\rm PBH})$ plane (see, for instance, Fig.~\ref{fig:SNR}). This region is further constrained once we impose a meaningful estimation of errors, quantified by the relative uncertainties $\delta \theta_i/\theta_i \leq 1$, obtained using the Fisher forecast formalism (see Fig.~\ref{fig:Fisher}). As the relative error threshold is tightened to $\delta \theta_i/\theta_i \leq 1$, the allowed PBH parameter space progressively shrinks, reflecting increasingly stringent constraints on parameter estimation. Quantitatively, the region satisfying $\mathrm{SNR} \geq 1$ corresponds to $M_{\rm PBH} \in (0.003,\,100)\,M_\odot$ and $f_{\rm PBH} \in (10^{-4.5},\,10^{-2.4})$. Imposing the additional conditions $\delta M_{\rm PBH}/M_{\rm PBH} \leq 1$ and $\delta f_{\rm PBH}/f_{\rm PBH} \leq 1$ further restricts the parameter space (see Fig.~\ref{fig:Fisher}), yielding $M_{\rm PBH} \in (0.02,\,100)\,M_\odot$, with a correspondingly narrower range in $f_{\rm PBH}$.

After quantifying the uncertainties in the PBH parameters, we propagate them to estimate the resulting uncertainty in $H_0$. The constraints are presented in  Fig.~\ref{fig:dH0contour}, and summarized in Table~\ref{tab:H0_results}. We find that the achievable precision on $H_0$ is strongly correlated with the accuracy of PBH parameters. In particular, for relative uncertainties $\delta \theta_i/\theta_i \leq 0.1$, the corresponding uncertainty in $H_0$ is $\delta H_0 \lesssim 2~{\rm km\,s^{-1}\,Mpc^{-1}}$, in a conservative approach. However, for an optimistic approach of precision measurement, by improving the parameter precision to $\delta \theta_i/\theta_i \leq 0.01$ reduces the uncertainty to $\delta H_0 \sim \mathcal{O}(0.1)$~${\rm km\,s^{-1}\,Mpc^{-1}}$ (see, right panel of  Fig.~\ref{fig:dH0contour}).

We have also examined the robustness of these results against variations in key assumptions. As demonstrated in Fig.~\ref{fig:H0_fiducials}, the relative uncertainty $\delta H_0/H_0$ is largely insensitive to the choice of the fiducial value of $H_0$ across the currently allowed range. In contrast, the collapse efficiency parameter $\gamma$ introduces a moderate dependence by modifying the mapping between PBH parameters and GW observables, leading to an $\mathcal{O}(10)$ variation in $\delta H_0$ (see Fig.~\ref{fig:dH0_gamma}). Nevertheless, this does not qualitatively alter the overall conclusions.

Overall, our analysis demonstrates that multi-band GW observations provide a promising and independent approach to constrain the uncertainty in $H_0$. The combination of induced and merger GW signals helps to break parameter degeneracies and enables the extraction of cosmological information directly from the GW spectrum. As illustrated in Figs.~\ref{fig:SNR}–\ref{fig:dH0_gamma} and Table~\ref{tab:H0_results}, this approach offers a complementary probe that is largely independent of the assumptions underlying traditional cosmological measurements. With the expected sensitivity of future GW detectors, such multi-band analyses have the potential to improve the precision of cosmological parameters and provide a consistency test of the standard cosmological model. The achieved precision demonstrates that multi-band GW observations can serve as a robust and independent probe of the Hubble parameter, without relying only on the cosmic distance ladder.

\section*{Acknowledgments}
We thank Debarun Paul and Rahul Shah for fruitful discussions. S thanks ISI, Kolkata for financial support through a Junior Research Fellowship. MRH acknowledges the Tsung-Dao Lee Institute, Shanghai Jiao Tong University, for financial support through the Siyuan Postdoctoral Fellowship.  SP thanks ANRF,  Govt. of India, for partial support through Project No. CRG/2023/003984. We acknowledge the computational facilities provided by the SyMeC HPC cluster of ISI Kolkata.
\bibliographystyle{JHEP}
\bibliography{Ref} 

\providecommand{\href}[2]{#2}\begingroup\raggedright\begin{thebibliography}{10}

\bibitem{LIGOScientific:2017vwq}
{\scshape LIGO Scientific, Virgo} collaboration, \emph{{GW170817: Observation of Gravitational Waves from a Binary Neutron Star Inspiral}}, \href{https://doi.org/10.1103/PhysRevLett.119.161101}{\emph{Phys. Rev. Lett.} {\bfseries 119} (2017) 161101} [\href{https://arxiv.org/abs/1710.05832}{{\ttfamily 1710.05832}}].

\bibitem{LIGOScientific:2017adf}
{\scshape LIGO Scientific, Virgo, 1M2H, Dark Energy Camera GW-E, DES, DLT40, Las Cumbres Observatory, VINROUGE, MASTER} collaboration, \emph{{A gravitational-wave standard siren measurement of the Hubble constant}}, \href{https://doi.org/10.1038/nature24471}{\emph{Nature} {\bfseries 551} (2017) 85} [\href{https://arxiv.org/abs/1710.05835}{{\ttfamily 1710.05835}}].

\bibitem{Schutz:1986gp}
B.F.~Schutz, \emph{{Determining the Hubble Constant from Gravitational Wave Observations}}, \href{https://doi.org/10.1038/323310a0}{\emph{Nature} {\bfseries 323} (1986) 310}.

\bibitem{LISACosmologyWorkingGroup:2025vdz}
{\scshape LISA Cosmology Working Group} collaboration, \emph{{Reconstructing primordial curvature perturbations via scalar-induced gravitational waves with LISA}}, \href{https://doi.org/10.1088/1475-7516/2025/05/062}{\emph{JCAP} {\bfseries 05} (2025) 062} [\href{https://arxiv.org/abs/2501.11320}{{\ttfamily 2501.11320}}].

\bibitem{LIGOScientific:2016aoc}
{\scshape LIGO Scientific, Virgo} collaboration, \emph{{Observation of Gravitational Waves from a Binary Black Hole Merger}}, \href{https://doi.org/10.1103/PhysRevLett.116.061102}{\emph{Phys. Rev. Lett.} {\bfseries 116} (2016) 061102} [\href{https://arxiv.org/abs/1602.03837}{{\ttfamily 1602.03837}}].

\bibitem{Abramovici:1992ah}
A.~Abramovici et~al., \emph{{LIGO: The Laser interferometer gravitational wave observatory}}, \href{https://doi.org/10.1126/science.256.5055.325}{\emph{Science} {\bfseries 256} (1992) 325}.

\bibitem{Abbott:1984fp}
L.F.~Abbott and M.B.~Wise, \emph{{Constraints on Generalized Inflationary Cosmologies}}, \href{https://doi.org/10.1016/0550-3213(84)90329-8}{\emph{Nucl. Phys. B} {\bfseries 244} (1984) 541}.

\bibitem{Allen:1987bk}
B.~Allen, \emph{{The Stochastic Gravity Wave Background in Cosmological Space-Times}}, \href{https://doi.org/10.1103/PhysRevD.37.2078}{\emph{Phys. Rev. D} {\bfseries 37} (1988) 2078}.

\bibitem{NANOGrav:2023gor}
{\scshape NANOGrav} collaboration, \emph{{The NANOGrav 15 yr Data Set: Evidence for a Gravitational-wave Background}}, \href{https://doi.org/10.3847/2041-8213/acdac6}{\emph{Astrophys. J. Lett.} {\bfseries 951} (2023) L8} [\href{https://arxiv.org/abs/2306.16213}{{\ttfamily 2306.16213}}].

\bibitem{Antoniadis:2023rey}
{\scshape EPTA, InPTA} collaboration, \emph{{The second data release from the European Pulsar Timing Array - III. Search for gravitational wave signals}}, \href{https://doi.org/10.1051/0004-6361/202346844}{\emph{Astron. Astrophys.} {\bfseries 678} (2023) A3} [\href{https://arxiv.org/abs/2306.16214}{{\ttfamily 2306.16214}}].

\bibitem{Reardon:2023gzh}
D.J.~Reardon et~al., \emph{{Search for an Isotropic Gravitational-wave Background with the Parkes Pulsar Timing Array}}, \href{https://doi.org/10.3847/2041-8213/acdd02}{\emph{Astrophys. J. Lett.} {\bfseries 951} (2023) L6} [\href{https://arxiv.org/abs/2306.16215}{{\ttfamily 2306.16215}}].

\bibitem{Xu:2023wog}
H.~Xu et~al., \emph{{Searching for the Nano-Hertz Stochastic Gravitational Wave Background with the Chinese Pulsar Timing Array Data Release I}}, \href{https://doi.org/10.1088/1674-4527/acdfa5}{\emph{Res. Astron. Astrophys.} {\bfseries 23} (2023) 075024} [\href{https://arxiv.org/abs/2306.16216}{{\ttfamily 2306.16216}}].

\bibitem{Zeldovich:1967lct}
Y.B.~Zel'dovich and I.D.~Novikov, \emph{{The Hypothesis of Cores Retarded during Expansion and the Hot Cosmological Model}}, {\emph{Sov. Astron.} {\bfseries 10} (1967) 602}.

\bibitem{Hawking:1971ei}
S.~Hawking, \emph{{Gravitationally collapsed objects of very low mass}}, \href{https://doi.org/10.1093/mnras/152.1.75}{\emph{Mon. Not. Roy. Astron. Soc.} {\bfseries 152} (1971) 75}.

\bibitem{Carr:1974nx}
B.J.~Carr and S.W.~Hawking, \emph{{Black holes in the early Universe}}, \href{https://doi.org/10.1093/mnras/168.2.399}{\emph{Mon. Not. Roy. Astron. Soc.} {\bfseries 168} (1974) 399}.

\bibitem{Carr:2020xqk}
B.~Carr and F.~Kuhnel, \emph{{Primordial Black Holes as Dark Matter: Recent Developments}}, \href{https://doi.org/10.1146/annurev-nucl-050520-125911}{\emph{Ann. Rev. Nucl. Part. Sci.} {\bfseries 70} (2020) 355} [\href{https://arxiv.org/abs/2006.02838}{{\ttfamily 2006.02838}}].

\bibitem{Green:2020jor}
A.M.~Green and B.J.~Kavanagh, \emph{{Primordial Black Holes as a dark matter candidate}}, \href{https://doi.org/10.1088/1361-6471/abc534}{\emph{J. Phys. G} {\bfseries 48} (2021) 043001} [\href{https://arxiv.org/abs/2007.10722}{{\ttfamily 2007.10722}}].

\bibitem{Carr:2021bzv}
B.~Carr and F.~Kuhnel, \emph{{Primordial black holes as dark matter candidates}}, \href{https://doi.org/10.21468/SciPostPhysLectNotes.48}{\emph{SciPost Phys. Lect. Notes} {\bfseries 48} (2022) 1} [\href{https://arxiv.org/abs/2110.02821}{{\ttfamily 2110.02821}}].

\bibitem{Bird:2016dcv}
S.~Bird, I.~Cholis, J.B.~Mu{\~n}oz, Y.~Ali-Ha{\"\i}moud, M.~Kamionkowski, E.D.~Kovetz et~al., \emph{{Did LIGO detect dark matter?}}, \href{https://doi.org/10.1103/PhysRevLett.116.201301}{\emph{Phys. Rev. Lett.} {\bfseries 116} (2016) 201301} [\href{https://arxiv.org/abs/1603.00464}{{\ttfamily 1603.00464}}].

\bibitem{Clesse:2016vqa}
S.~Clesse and J.~Garc{\'\i}a-Bellido, \emph{{The clustering of massive Primordial Black Holes as Dark Matter: measuring their mass distribution with Advanced LIGO}}, \href{https://doi.org/10.1016/j.dark.2016.10.002}{\emph{Phys. Dark Univ.} {\bfseries 15} (2017) 142} [\href{https://arxiv.org/abs/1603.05234}{{\ttfamily 1603.05234}}].

\bibitem{Sasaki:2016jop}
M.~Sasaki, T.~Suyama, T.~Tanaka and S.~Yokoyama, \emph{{Primordial Black Hole Scenario for the Gravitational-Wave Event GW150914}}, \href{https://doi.org/10.1103/PhysRevLett.117.061101}{\emph{Phys. Rev. Lett.} {\bfseries 117} (2016) 061101} [\href{https://arxiv.org/abs/1603.08338}{{\ttfamily 1603.08338}}].

\bibitem{Carr:1975qj}
B.J.~Carr, \emph{{The Primordial black hole mass spectrum}}, \href{https://doi.org/10.1086/153853}{\emph{Astrophys. J.} {\bfseries 201} (1975) 1}.

\bibitem{Sasaki:2018dmp}
M.~Sasaki, T.~Suyama, T.~Tanaka and S.~Yokoyama, \emph{{Primordial black holes{\textemdash}perspectives in gravitational wave astronomy}}, \href{https://doi.org/10.1088/1361-6382/aaa7b4}{\emph{Class. Quant. Grav.} {\bfseries 35} (2018) 063001} [\href{https://arxiv.org/abs/1801.05235}{{\ttfamily 1801.05235}}].

\bibitem{Josan:2009qn}
A.S.~Josan, A.M.~Green and K.A.~Malik, \emph{{Generalised constraints on the primordial power spectrum from primordial black holes}}, \href{https://doi.org/10.1103/PhysRevD.79.103520}{\emph{Phys. Rev. D} {\bfseries 79} (2009) 103520} [\href{https://arxiv.org/abs/0903.3184}{{\ttfamily 0903.3184}}].

\bibitem{Matarrese:1997ay}
S.~Matarrese, S.~Mollerach and M.~Bruni, \emph{{Second-order perturbations of the Einstein-de Sitter universe}}, \href{https://doi.org/10.1103/PhysRevD.58.043504}{\emph{Phys. Rev. D} {\bfseries 58} (1998) 043504} [\href{https://arxiv.org/abs/astro-ph/9707278}{{\ttfamily astro-ph/9707278}}].

\bibitem{Ananda:2006af}
K.N.~Ananda, C.~Clarkson and D.~Wands, \emph{{The Cosmological gravitational wave background from primordial density perturbations}}, \href{https://doi.org/10.1103/PhysRevD.75.123518}{\emph{Phys. Rev. D} {\bfseries 75} (2007) 123518} [\href{https://arxiv.org/abs/gr-qc/0612013}{{\ttfamily gr-qc/0612013}}].

\bibitem{Baumann:2007zm}
D.~Baumann, P.J.~Steinhardt, K.~Takahashi and K.~Ichiki, \emph{{Gravitational Wave Spectrum Induced by Primordial Scalar Perturbations}}, \href{https://doi.org/10.1103/PhysRevD.76.084019}{\emph{Phys. Rev. D} {\bfseries 76} (2007) 084019} [\href{https://arxiv.org/abs/hep-th/0703290}{{\ttfamily hep-th/0703290}}].

\bibitem{DiValentino:2021izs}
E.~Di~Valentino, O.~Mena, S.~Pan, L.~Visinelli, W.~Yang, A.~Melchiorri et~al., \emph{{In the realm of the Hubble tension{\textemdash}a review of solutions}}, \href{https://doi.org/10.1088/1361-6382/ac086d}{\emph{Class. Quant. Grav.} {\bfseries 38} (2021) 153001} [\href{https://arxiv.org/abs/2103.01183}{{\ttfamily 2103.01183}}].

\bibitem{Abdalla:2022yfr}
E.~Abdalla et~al., \emph{{Cosmology intertwined: A review of the particle physics, astrophysics, and cosmology associated with the cosmological tensions and anomalies}}, \href{https://doi.org/10.1016/j.jheap.2022.04.002}{\emph{JHEAp} {\bfseries 34} (2022) 49} [\href{https://arxiv.org/abs/2203.06142}{{\ttfamily 2203.06142}}].

\bibitem{Planck:2018vyc}
{\scshape Planck} collaboration, \emph{{Planck 2018 results. VI. Cosmological parameters}}, \href{https://doi.org/10.1051/0004-6361/201833910}{\emph{Astron. Astrophys.} {\bfseries 641} (2020) A6} [\href{https://arxiv.org/abs/1807.06209}{{\ttfamily 1807.06209}}].

\bibitem{eBOSS:2020yzd}
{\scshape eBOSS} collaboration, \emph{{Completed SDSS-IV extended Baryon Oscillation Spectroscopic Survey: Cosmological implications from two decades of spectroscopic surveys at the Apache Point Observatory}}, \href{https://doi.org/10.1103/PhysRevD.103.083533}{\emph{Phys. Rev. D} {\bfseries 103} (2021) 083533} [\href{https://arxiv.org/abs/2007.08991}{{\ttfamily 2007.08991}}].

\bibitem{Riess:2021jrx}
A.G.~Riess et~al., \emph{{A Comprehensive Measurement of the Local Value of the Hubble Constant with 1 km s$^{−1}$ Mpc$^{−1}$ Uncertainty from the Hubble Space Telescope and the SH0ES Team}}, \href{https://doi.org/10.3847/2041-8213/ac5c5b}{\emph{Astrophys. J. Lett.} {\bfseries 934} (2022) L7} [\href{https://arxiv.org/abs/2112.04510}{{\ttfamily 2112.04510}}].

\bibitem{Liu:2021jnw}
L.~Liu, X.-Y.~Yang, Z.-K.~Guo and R.-G.~Cai, \emph{{Testing primordial black hole and measuring the Hubble constant with multiband gravitational-wave observations}}, \href{https://doi.org/10.1088/1475-7516/2023/01/006}{\emph{JCAP} {\bfseries 01} (2023) 006} [\href{https://arxiv.org/abs/2112.05473}{{\ttfamily 2112.05473}}].

\bibitem{Kohri:2018awv}
K.~Kohri and T.~Terada, \emph{{Semianalytic calculation of gravitational wave spectrum nonlinearly induced from primordial curvature perturbations}}, \href{https://doi.org/10.1103/PhysRevD.97.123532}{\emph{Phys. Rev. D} {\bfseries 97} (2018) 123532} [\href{https://arxiv.org/abs/1804.08577}{{\ttfamily 1804.08577}}].

\bibitem{Saito:2008jc}
R.~Saito and J.~Yokoyama, \emph{{Gravitational wave background as a probe of the primordial black hole abundance}}, \href{https://doi.org/10.1103/PhysRevLett.102.161101}{\emph{Phys. Rev. Lett.} {\bfseries 102} (2009) 161101} [\href{https://arxiv.org/abs/0812.4339}{{\ttfamily 0812.4339}}].

\bibitem{Dom_nech_2020}
G.~Domènech, \emph{Induced gravitational waves in a general cosmological background}, \href{https://doi.org/10.1142/s0218271820500285}{\emph{International Journal of Modern Physics D} {\bfseries 29} (2020) 2050028}.

\bibitem{Domenech:2021ztg}
G.~Dom{\`e}nech, \emph{{Scalar Induced Gravitational Waves Review}}, \href{https://doi.org/10.3390/universe7110398}{\emph{Universe} {\bfseries 7} (2021) 398} [\href{https://arxiv.org/abs/2109.01398}{{\ttfamily 2109.01398}}].

\bibitem{Vieira:2026eof}
N.~Vieira et~al., \emph{{Search For a Counterpart to the Subsolar Mass Gravitational Wave Candidate S251112cm}},  \href{https://arxiv.org/abs/2603.17009}{{\ttfamily 2603.17009}}.

\bibitem{Haque:2026yum}
M.R.~Haque, F.~Iocco and L.~Visinelli, \emph{{Primordial Black Hole interpretation of the sub-solar merger event S251112cm}},  \href{https://arxiv.org/abs/2603.25795}{{\ttfamily 2603.25795}}.

\bibitem{Alabidi:2012ex}
L.~Alabidi, K.~Kohri, M.~Sasaki and Y.~Sendouda, \emph{{Observable Spectra of Induced Gravitational Waves from Inflation}}, \href{https://doi.org/10.1088/1475-7516/2012/09/017}{\emph{JCAP} {\bfseries 09} (2012) 017} [\href{https://arxiv.org/abs/1203.4663}{{\ttfamily 1203.4663}}].

\bibitem{Nakama:2016enz}
T.~Nakama and T.~Suyama, \emph{{Primordial black holes as a novel probe of primordial gravitational waves. II: Detailed analysis}}, \href{https://doi.org/10.1103/PhysRevD.94.043507}{\emph{Phys. Rev. D} {\bfseries 94} (2016) 043507} [\href{https://arxiv.org/abs/1605.04482}{{\ttfamily 1605.04482}}].

\bibitem{NANOGrav:2023hvm}
{\scshape NANOGrav} collaboration, \emph{{The NANOGrav 15 yr Data Set: Search for Signals from New Physics}}, \href{https://doi.org/10.3847/2041-8213/acdc91}{\emph{Astrophys. J. Lett.} {\bfseries 951} (2023) L11} [\href{https://arxiv.org/abs/2306.16219}{{\ttfamily 2306.16219}}].

\bibitem{Janssen:2015dca}
G.~Janssen et~al., \emph{{Gravitational wave astronomy with the SKA}}, \href{https://doi.org/10.22323/1.215.0037}{\emph{PoS} {\bfseries AASKA14} (2015) 037} [\href{https://arxiv.org/abs/1501.00127}{{\ttfamily 1501.00127}}].

\bibitem{Weltman:2019neo}
A.~Weltman et~al., \emph{{Fundamental physics with the Square Kilometre Array}}, \href{https://doi.org/10.1017/pasa.2019.42}{\emph{Publ. Astron. Soc. Austral.} {\bfseries 37} (2020) e002} [\href{https://arxiv.org/abs/2002.11733}{{\ttfamily 2002.11733}}].

\bibitem{Wang:2016ana}
S.~Wang, Y.-F.~Wang, Q.-G.~Huang and T.G.F.~Li, \emph{{Constraints on the Primordial Black Hole Abundance from the First Advanced LIGO Observation Run Using the Stochastic Gravitational-Wave Background}}, \href{https://doi.org/10.1103/PhysRevLett.120.191102}{\emph{Phys. Rev. Lett.} {\bfseries 120} (2018) 191102} [\href{https://arxiv.org/abs/1610.08725}{{\ttfamily 1610.08725}}].

\bibitem{Raidal:2017mfl}
M.~Raidal, V.~Vaskonen and H.~Veerm{\"a}e, \emph{{Gravitational Waves from Primordial Black Hole Mergers}}, \href{https://doi.org/10.1088/1475-7516/2017/09/037}{\emph{JCAP} {\bfseries 09} (2017) 037} [\href{https://arxiv.org/abs/1707.01480}{{\ttfamily 1707.01480}}].

\bibitem{KAGRA:2021kbb}
{\scshape KAGRA, Virgo, LIGO Scientific} collaboration, \emph{{Upper limits on the isotropic gravitational-wave background from Advanced LIGO and Advanced Virgo\u2019s third observing run}}, \href{https://doi.org/10.1103/PhysRevD.104.022004}{\emph{Phys. Rev. D} {\bfseries 104} (2021) 022004} [\href{https://arxiv.org/abs/2101.12130}{{\ttfamily 2101.12130}}].

\bibitem{KAGRA:2025iso}
{\scshape LIGO Scientific, Virgo, KAGRA} collaboration, \emph{{Upper Limits on the Isotropic Gravitational-Wave Background from the first part of LIGO, Virgo, and KAGRA's fourth Observing Run}},  \href{https://arxiv.org/abs/2508.20721}{{\ttfamily 2508.20721}}.

\bibitem{KAGRA:2025cosmo}
{\scshape LIGO Scientific, Virgo, KAGRA} collaboration, \emph{{Cosmological and High Energy Physics implications from gravitational-wave background searches in LIGO-Virgo-KAGRA's O1-O4a runs}},  \href{https://arxiv.org/abs/2510.26848}{{\ttfamily 2510.26848}}.

\bibitem{Punturo:2010zz}
M.~Punturo et~al., \emph{{The Einstein Telescope: A third-generation gravitational wave observatory}}, \href{https://doi.org/10.1088/0264-9381/27/19/194002}{\emph{Class. Quant. Grav.} {\bfseries 27} (2010) 194002}.

\bibitem{Maggiore:2019uih}
M.~Maggiore et~al., \emph{{Science Case for the Einstein Telescope}}, \href{https://doi.org/10.1088/1475-7516/2020/03/050}{\emph{JCAP} {\bfseries 03} (2020) 050} [\href{https://arxiv.org/abs/1912.02622}{{\ttfamily 1912.02622}}].

\bibitem{Allen:1996vm}
B.~Allen, \emph{{The Stochastic gravity wave background: Sources and detection}},  \href{https://arxiv.org/abs/gr-qc/9604033}{{\ttfamily gr-qc/9604033}}.

\bibitem{Allen:1997ad}
B.~Allen and J.D.~Romano, \emph{{Detecting a stochastic background of gravitational radiation: Signal processing strategies and sensitivities}}, \href{https://doi.org/10.1103/PhysRevD.59.102001}{\emph{Phys. Rev. D} {\bfseries 59} (1999) 102001} [\href{https://arxiv.org/abs/gr-qc/9710117}{{\ttfamily gr-qc/9710117}}].

\bibitem{Flanagan:1997kp}
E.E.~Flanagan and S.A.~Hughes, \emph{{The Basics of gravitational wave theory}}, \href{https://doi.org/10.1103/PhysRevD.57.4535}{\emph{Phys. Rev. D} {\bfseries 57} (1998) 4535} [\href{https://arxiv.org/abs/gr-qc/9701039}{{\ttfamily gr-qc/9701039}}].

\bibitem{Cutler:1994ys}
C.~Cutler and E.E.~Flanagan, \emph{{Gravitational waves from merging compact binaries: How accurately can one extract the binary's parameters from the inspiral waveform?}}, \href{https://doi.org/10.1103/PhysRevD.49.2658}{\emph{Phys. Rev. D} {\bfseries 49} (1994) 2658} [\href{https://arxiv.org/abs/gr-qc/9402014}{{\ttfamily gr-qc/9402014}}].

\bibitem{Poisson:1995ef}
E.~Poisson and C.M.~Will, \emph{{Gravitational waves from inspiraling compact binaries: Parameter estimation using second postNewtonian asymptotics}}, \href{https://doi.org/10.1103/PhysRevD.52.848}{\emph{Phys. Rev. D} {\bfseries 52} (1995) 848} [\href{https://arxiv.org/abs/gr-qc/9502040}{{\ttfamily gr-qc/9502040}}].

\bibitem{Musco:2012au}
I.~Musco and J.C.~Miller, \emph{{Primordial black hole formation in the early universe: critical behaviour and self-similarity}}, \href{https://doi.org/10.1088/0264-9381/30/14/145009}{\emph{Class. Quant. Grav.} {\bfseries 30} (2013) 145009} [\href{https://arxiv.org/abs/1201.2379}{{\ttfamily 1201.2379}}].

\bibitem{Musco:2008hv}
I.~Musco, J.C.~Miller and A.G.~Polnarev, \emph{{Primordial black hole formation in the radiative era: Investigation of the critical nature of the collapse}}, \href{https://doi.org/10.1088/0264-9381/26/23/235001}{\emph{Class. Quant. Grav.} {\bfseries 26} (2009) 235001} [\href{https://arxiv.org/abs/0811.1452}{{\ttfamily 0811.1452}}].

\bibitem{Hawke:2002rf}
I.~Hawke and J.M.~Stewart, \emph{{The dynamics of primordial black hole formation}}, \href{https://doi.org/10.1088/0264-9381/19/14/310}{\emph{Class. Quant. Grav.} {\bfseries 19} (2002) 3687}.

\bibitem{Niemeyer:1997mt}
J.C.~Niemeyer and K.~Jedamzik, \emph{{Near-critical gravitational collapse and the initial mass function of primordial black holes}}, \href{https://doi.org/10.1103/PhysRevLett.80.5481}{\emph{Phys. Rev. Lett.} {\bfseries 80} (1998) 5481} [\href{https://arxiv.org/abs/astro-ph/9709072}{{\ttfamily astro-ph/9709072}}].

\bibitem{Escriva:2021pmf}
A.~Escriv\`a and A.E.~Romano, \emph{{Effects of the shape of curvature peaks on the size of primordial black holes}}, \href{https://doi.org/10.1088/1475-7516/2021/05/066}{\emph{JCAP} {\bfseries 05} (2021) 066} [\href{https://arxiv.org/abs/2103.03867}{{\ttfamily 2103.03867}}].

\bibitem{Escriva:2019nsa}
A.~Escriv\`a, \emph{{Simulation of primordial black hole formation using pseudo-spectral methods}}, \href{https://doi.org/10.1016/j.dark.2020.100466}{\emph{Phys. Dark Univ.} {\bfseries 27} (2020) 100466} [\href{https://arxiv.org/abs/1907.13065}{{\ttfamily 1907.13065}}].

\bibitem{Escriva:2020tak}
A.~Escriv{\`a}, C.~Germani and R.K.~Sheth, \emph{{Analytical thresholds for black hole formation in general cosmological backgrounds}}, \href{https://doi.org/10.1088/1475-7516/2021/01/030}{\emph{JCAP} {\bfseries 01} (2021) 030} [\href{https://arxiv.org/abs/2007.05564}{{\ttfamily 2007.05564}}].

\bibitem{Escriva:2021aeh}
A.~Escriv\`a, \emph{{PBH Formation from Spherically Symmetric Hydrodynamical Perturbations: A Review}}, \href{https://doi.org/10.3390/universe8020066}{\emph{Universe} {\bfseries 8} (2022) 66} [\href{https://arxiv.org/abs/2111.12693}{{\ttfamily 2111.12693}}].

\bibitem{Wands:2000dp}
D.~Wands, K.A.~Malik, D.H.~Lyth and A.R.~Liddle, \emph{{A New approach to the evolution of cosmological perturbations on large scales}}, \href{https://doi.org/10.1103/PhysRevD.62.043527}{\emph{Phys. Rev. D} {\bfseries 62} (2000) 043527} [\href{https://arxiv.org/abs/astro-ph/0003278}{{\ttfamily astro-ph/0003278}}].

\bibitem{Harada:2013epa}
T.~Harada, C.-M.~Yoo and K.~Kohri, \emph{{Threshold of primordial black hole formation}}, \href{https://doi.org/10.1103/PhysRevD.88.084051}{\emph{Phys. Rev. D} {\bfseries 88} (2013) 084051} [\href{https://arxiv.org/abs/1309.4201}{{\ttfamily 1309.4201}}].

\bibitem{Ajith:2007kx}
P.~Ajith et~al., \emph{{A Template bank for gravitational waveforms from coalescing binary black holes. I. Non-spinning binaries}}, \href{https://doi.org/10.1103/PhysRevD.77.104017}{\emph{Phys. Rev. D} {\bfseries 77} (2008) 104017} [\href{https://arxiv.org/abs/0710.2335}{{\ttfamily 0710.2335}}].

\bibitem{Ajith:2009bn}
P.~Ajith et~al., \emph{{Inspiral-merger-ringdown waveforms for black-hole binaries with non-precessing spins}}, \href{https://doi.org/10.1103/PhysRevLett.106.241101}{\emph{Phys. Rev. Lett.} {\bfseries 106} (2011) 241101} [\href{https://arxiv.org/abs/0909.2867}{{\ttfamily 0909.2867}}].

\bibitem{Raidal:2018bbj}
M.~Raidal, C.~Spethmann, V.~Vaskonen and H.~Veerm{\"a}e, \emph{{Formation and Evolution of Primordial Black Hole Binaries in the Early Universe}}, \href{https://doi.org/10.1088/1475-7516/2019/02/018}{\emph{JCAP} {\bfseries 02} (2019) 018} [\href{https://arxiv.org/abs/1812.01930}{{\ttfamily 1812.01930}}].

\bibitem{Liu:2018ess}
L.~Liu, Z.-K.~Guo and R.-G.~Cai, \emph{{Effects of the surrounding primordial black holes on the merger rate of primordial black hole binaries}}, \href{https://doi.org/10.1103/PhysRevD.99.063523}{\emph{Phys. Rev. D} {\bfseries 99} (2019) 063523} [\href{https://arxiv.org/abs/1812.05376}{{\ttfamily 1812.05376}}].

\bibitem{Vaskonen:2019jpv}
V.~Vaskonen and H.~Veerm{\"a}e, \emph{{Lower bound on the primordial black hole merger rate}}, \href{https://doi.org/10.1103/PhysRevD.101.043015}{\emph{Phys. Rev. D} {\bfseries 101} (2020) 043015} [\href{https://arxiv.org/abs/1908.09752}{{\ttfamily 1908.09752}}].

\bibitem{Hutsi:2020sol}
G.~H{\"u}tsi, M.~Raidal, V.~Vaskonen and H.~Veerm{\"a}e, \emph{{Two populations of LIGO-Virgo black holes}}, \href{https://doi.org/10.1088/1475-7516/2021/03/068}{\emph{JCAP} {\bfseries 03} (2021) 068} [\href{https://arxiv.org/abs/2012.02786}{{\ttfamily 2012.02786}}].

\bibitem{Baumann:2009ds}
D.~Baumann, \emph{{Inflation}},  in \emph{{Theoretical Advanced Study Institute in Elementary Particle Physics}: {Physics of the Large and the Small}}, pp.~523--686, 2011, \href{https://doi.org/10.1142/9789814327183_0010}{DOI} [\href{https://arxiv.org/abs/0907.5424}{{\ttfamily 0907.5424}}].

\bibitem{Cai:2018dig}
R.-g.~Cai, S.~Pi and M.~Sasaki, \emph{{Gravitational Waves Induced by non-Gaussian Scalar Perturbations}}, \href{https://doi.org/10.1103/PhysRevLett.122.201101}{\emph{Phys. Rev. Lett.} {\bfseries 122} (2019) 201101} [\href{https://arxiv.org/abs/1810.11000}{{\ttfamily 1810.11000}}].

\bibitem{Thrane:2013oya}
E.~Thrane and J.D.~Romano, \emph{{Sensitivity curves for searches for gravitational-wave backgrounds}}, \href{https://doi.org/10.1103/PhysRevD.88.124032}{\emph{Phys. Rev. D} {\bfseries 88} (2013) 124032} [\href{https://arxiv.org/abs/1310.5300}{{\ttfamily 1310.5300}}].

\bibitem{Caprini:2015zlo}
C.~Caprini et~al., \emph{{Science with the space-based interferometer eLISA. II: Gravitational waves from cosmological phase transitions}}, \href{https://doi.org/10.1088/1475-7516/2016/04/001}{\emph{JCAP} {\bfseries 04} (2016) 001} [\href{https://arxiv.org/abs/1512.06239}{{\ttfamily 1512.06239}}].

\bibitem{Schmitz:2020syl}
K.~Schmitz, \emph{{New Sensitivity Curves for Gravitational-Wave Signals from Cosmological Phase Transitions}}, \href{https://doi.org/10.1007/JHEP01(2021)097}{\emph{JHEP} {\bfseries 01} (2021) 097} [\href{https://arxiv.org/abs/2002.04615}{{\ttfamily 2002.04615}}].

\bibitem{Gowling:2021gcy}
C.~Gowling and M.~Hindmarsh, \emph{{Observational prospects for phase transitions at LISA: Fisher matrix analysis}}, \href{https://doi.org/10.1088/1475-7516/2021/10/039}{\emph{JCAP} {\bfseries 10} (2021) 039} [\href{https://arxiv.org/abs/2106.05984}{{\ttfamily 2106.05984}}].

\bibitem{Dodelson:2003ft}
S.~Dodelson, \emph{{Modern Cosmology}}, Academic Press, Amsterdam (2003).

\bibitem{Carr:2026hot}
B.~Carr, A.J.~Iovino, G.~Perna, V.~Vaskonen and H.~Veerm{\"a}e, \emph{{Primordial black holes: constraints, potential evidence and prospects}},  \href{https://arxiv.org/abs/2601.06024}{{\ttfamily 2601.06024}}.

\bibitem{Oncins:2022ydg}
M.~Oncins, \emph{{Constraints on PBH as dark matter from observations: a review}},  \href{https://arxiv.org/abs/2205.14722}{{\ttfamily 2205.14722}}.

\bibitem{Carr:2016drx}
B.~Carr, F.~Kuhnel and M.~Sandstad, \emph{{Primordial Black Holes as Dark Matter}}, \href{https://doi.org/10.1103/PhysRevD.94.083504}{\emph{Phys. Rev. D} {\bfseries 94} (2016) 083504} [\href{https://arxiv.org/abs/1607.06077}{{\ttfamily 1607.06077}}].

\bibitem{Niikura:2019kqi}
H.~Niikura, M.~Takada, S.~Yokoyama, T.~Sumi and S.~Masaki, \emph{{Constraints on Earth-mass primordial black holes from OGLE 5-year microlensing events}}, \href{https://doi.org/10.1103/PhysRevD.99.083503}{\emph{Phys. Rev. D} {\bfseries 99} (2019) 083503} [\href{https://arxiv.org/abs/1901.07120}{{\ttfamily 1901.07120}}].

\bibitem{Niikura:2017zjd}
H.~Niikura et~al., \emph{{Microlensing constraints on primordial black holes with Subaru/HSC Andromeda observations}}, \href{https://doi.org/10.1038/s41550-019-0723-1}{\emph{Nature Astron.} {\bfseries 3} (2019) 524} [\href{https://arxiv.org/abs/1701.02151}{{\ttfamily 1701.02151}}].

\bibitem{Mroz:2024mse}
P.~Mr{\'o}z et~al., \emph{{No massive black holes in the Milky Way halo}}, \href{https://doi.org/10.1038/s41586-024-07704-6}{\emph{Nature} {\bfseries 632} (2024) 749} [\href{https://arxiv.org/abs/2403.02386}{{\ttfamily 2403.02386}}].

\bibitem{Nitz:2021vqh}
A.H.~Nitz and Y.-F.~Wang, \emph{{Search for Gravitational Waves from the Coalescence of Subsolar-Mass Binaries in the First Half of Advanced LIGO and Virgo{\textquoteright}s Third Observing Run}}, \href{https://doi.org/10.1103/PhysRevLett.127.151101}{\emph{Phys. Rev. Lett.} {\bfseries 127} (2021) 151101} [\href{https://arxiv.org/abs/2106.08979}{{\ttfamily 2106.08979}}].

\bibitem{Kavanagh:2018ggo}
B.J.~Kavanagh, D.~Gaggero and G.~Bertone, \emph{{Merger rate of a subdominant population of primordial black holes}}, \href{https://doi.org/10.1103/PhysRevD.98.023536}{\emph{Phys. Rev. D} {\bfseries 98} (2018) 023536} [\href{https://arxiv.org/abs/1805.09034}{{\ttfamily 1805.09034}}].

\bibitem{Planck:2018vyg}
{\scshape Planck} collaboration, \emph{{Planck 2018 results. VI. Cosmological parameters}}, \href{https://doi.org/10.1051/0004-6361/201833910}{\emph{Astron. Astrophys.} {\bfseries 641} (2020) A6} [\href{https://arxiv.org/abs/1807.06209}{{\ttfamily 1807.06209}}].

\bibitem{H0DN:2025lyy}
{\scshape H0DN} collaboration, \emph{{The Local Distance Network: a community consensus report on the measurement of the Hubble constant at 1{\%} precision}},  \href{https://arxiv.org/abs/2510.23823}{{\ttfamily 2510.23823}}.

\bibitem{Carr:2009jm}
B.J.~Carr, K.~Kohri, Y.~Sendouda and J.~Yokoyama, \emph{{New cosmological constraints on primordial black holes}}, \href{https://doi.org/10.1103/PhysRevD.81.104019}{\emph{Phys. Rev. D} {\bfseries 81} (2010) 104019} [\href{https://arxiv.org/abs/0912.5297}{{\ttfamily 0912.5297}}].

\end{thebibliography}\endgroup
\end{document}